\documentclass[sn-mathphys-num,iicol]{sn-jnl}

\usepackage{graphicx}%
\usepackage{multirow}%
\usepackage{amsmath,amssymb,amsfonts}%
\usepackage{amsthm}%
\usepackage{mathrsfs}%
\usepackage[title]{appendix}%
\usepackage{xcolor}%
\usepackage{textcomp}%
\usepackage{manyfoot}%
\usepackage{booktabs}%
\usepackage{algorithm}%
\usepackage{algorithmicx}%
\usepackage{algpseudocode}%
\usepackage{listings}%
\usepackage{tikz, pgfplots}
\usepackage{pslatex}

\usepackage{lmodern}
\usepackage{anyfontsize}
\newcommand{\R}{\mathbb{R}}
\DeclareMathOperator{\prox}{prox}
\DeclareMathOperator{\DB}{DB}
\DeclareMathOperator{\Down}{Down}
\DeclareMathOperator{\BU}{BD^t}
\DeclareMathOperator{\Up}{Up}
\DeclareMathOperator{\HS}{H}
\DeclareMathOperator{\U}{U}
\DeclareMathOperator{\V}{V}
\DeclareMathOperator{\TT}{T}
\DeclareMathOperator{\PAN}{P}

\DeclareMathOperator{\MARNet}{MARNet}
\DeclareMathOperator{\argmin}{argmin}
\DeclareMathOperator{\Bicubic}{Bicubic}

\begin{document}

\title[Multi-Head Attention Residual Unfolded Network for Model-Based Pansharpening]{Multi-Head Attention Residual Unfolded Network for Model-Based Pansharpening}


\author*[1,2]{\fnm{Ivan} \sur{Pereira-Sánchez}}\email{i.pereira@uib.es}

\author[1,2]{\fnm{Eloi} \sur{Sans}}\email{e.sans@uib.es}

\author[1,2]{\fnm{Julia} \sur{Navarro}}\email{julia.navarro@uib.es}
\author[1,2]{\fnm{Joan} \sur{Duran}}\email{joan.duran@uib.es}

\affil[1]{\orgdiv{Dept.~of Mathematics and Computer Science}, \orgname{Universitat de les Illess Balears (UIB)}, \orgaddress{\street{Cra.~de Valldemossa km 7.5}, \postcode{E-07122} \city{Palma}, \country{Spain}}}

\affil[2]{\orgdiv{Institute of Applied Computing and Community Code}, \orgname{UIB}, \orgaddress{\street{Edifici Complexe d’R+D, Cra.~de Valldemossa km 7.4}, \postcode{E-07122} \city{Palma}, \country{Spain}}}


\abstract{The objective of pansharpening and hypersharpening is to accurately combine a high-resolution panchromatic (PAN) image with a low-resolution multispectral (MS) or hyperspectral (HS) image, respectively. Unfolding fusion methods integrate the powerful representation capabilities of deep learning with the robustness of model-based approaches. These techniques involve unrolling the steps of the optimization scheme derived from the minimization of an energy into a deep learning framework, resulting in efficient and highly interpretable architectures. In this paper, we propose a model-based deep unfolded method for satellite image fusion. Our approach is based on a variational formulation that incorporates the classic observation model for MS/HS data, a high-frequency injection constraint based on the PAN image, and an arbitrary convex prior. For the unfolding stage, we introduce upsampling and downsampling layers that use geometric information encoded in the PAN image through residual networks. The backbone of our method is a multi-head attention residual network (MARNet), which replaces the proximity operator in the optimization scheme and combines multiple head attentions with residual learning to exploit image self-similarities via nonlocal operators defined in terms of patches. Additionally, we incorporate a post-processing module based on the MARNet architecture to further enhance the quality of the fused images. Experimental results on PRISMA, Quickbird, and WorldView2 datasets demonstrate the superior performance of our method and its ability to generalize across different sensor configurations and varying spatial and spectral resolutions. The source code will be available at \url{https://github.com/TAMI-UIB/MARNet}.}



\maketitle
\section{Introduction}

Earth observation satellites are useful for a wide variety of applications, including telecommunications, weather forecasting, navigation, environmental monitoring, cartography, and surveillance \cite{navalgund2007remote}. Many of these satellites, such as Ikonos, Landsat, WorldView, PRISMA, and Pléiades use passive remote sensing sensors that capture the reflection of waves produced by external sources \cite{ulaby1981microwave}.

Due to technical limitations related to on-board storage and bandwidth transmission, passive satellites are usually equipped with two types of sensors that have complementary attributes. On the one hand, the panchromatic (PAN) sensor generates a high-resolution image containing reflectance data across a broad range of wavelengths, which accurately represents the geometry of the scene. Conversely, the spectral sensors focus on narrower bandwidths, providing detailed information on the chemical-physical properties of objects. To compensate for the limited energy available in a specific spectral range, these sensors usually sample over a larger spatial area, resulting in lower resolution images. Depending on the number of bands, spectral sensors are called multispectral (MS) or hyperspectral (HS). MS sensors typically capture from 3 to 16 bands in the visible and infrared spectrum, while HS sensors capture the electromagnetic spectrum across dozens or even hundreds of bands.

Images with high spatial and spectral resolutions are essential for numerous applications. Consequently, extensive research has been conducted on satellite image fusion \cite{loncan2015hyperspectral, meng2019review, javan2021review, liangjiandeng2023pancollection}. The primary objective is to integrate the geometric accuracy from the PAN image with the spectral consistency of the MS/HS data, resulting in a single high-resolution MS/HS image. These fusion processes are referred to as pansharpening and hypersharpening, depending on the number of spectral bands involved.

\begin{table}[t]
    \centering
    \caption{List of abbreviations and notations used throughout this paper. 
    }
    
    \begin{tabular}{c|c}
        \hline
        Symbol & Concept\\ 
        \hline
        PAN &  Panchromatic\\
        MS & Multispectral\\
        HS & Hyperspectral\\
        $H$, $W$ & Height and width in high resolution\\
        $h$, $w$ &  Height and width in low resolution\\
        $C$ & Number of spectral bands\\
        $s$ & Sampling factor\\
        D & $s$-fold decimation operator\\
        B & Low-pass filtering operator \\ 
        $\PAN $ & PAN image \\
        $\widehat{\PAN}$ & Low-pass version of the PAN image\\
        $\U$ & Fused high-resolution MS/HS image \\
        $\HS$ & Low-resolution MS/HS image \\
        $\widehat{\HS}$ & Upsampled MS/HS image\\
         \hline
    \end{tabular}
    \label{tab:notation}
\end{table}

Classic fusion methods typically use component substitution \cite{gillespie1987color,chavez1991comparison,kwarteng1989extracting,aiazzi2007improving}, multi-resolution analysis \cite{nunez1999multiresolution,king2001wavelet,aiazzi2002context,lee2009fast}, or variational optimization \cite{ballester2006variational, duran2014nonlocal,fu2019variational}. These approaches are adaptable to changes in modeling assumptions due to their plug-and-play nature and do not require extensive training datasets, though they often rely on hand-crafted priors. On the other hand, recent advancements have introduced numerous deep learning methods, which can be broadly categorized into pure deep learning-based techniques \cite{Yang_2017_ICCV,deng2020detail,lu2023awfln} and model-based deep unfolding approaches \cite{xie2020mhf,xu2021deep,mai2024deep}. The first ones excel at learning natural priors, but tend to be less flexible and interpretable. Unfolding methods unroll the steps of the optimization scheme derived from the minimization of a model-based energy into a deep learning framework, resulting in efficient and interpretable architectures.

In this paper, we propose a model-based deep unfolded method to fuse a PAN image and a MS/HS image. We integrate residual networks with multi-head attention mechanisms to leverage image self-similarities through nonlocal operations. Our main contributions can be summarized as follows:
\begin{itemize}
    \item Based on \cite{duran2017survey}, we propose a variational model that incorporates an arbitrary convex regularization term, the classic observation model for MS/HS data \cite{ballester2006variational}, and a high-frequency injection constraint. In this regard, the most significant novelty is that this constraint is penalized using the $L^1$ norm, which better addresses noise correlation caused by interpolation. The resulting non-smooth energy is handled using the primal-dual Chambolle-Pock algorithm \cite{chambolle2011first}, which provides an optimization scheme that is subsequently unfolded.
    
    \item Inspired by residual \cite{he2016deep} and nonlocal networks \cite{wang2018non}, we replace the proximity operator in the primal-dual scheme with a novel nonlocal residual architecture that incorporates multiple head attentions. These attention mechanisms are specifically designed to capture spatial and spectral information by computing similarity weights across both PAN and MS/HS data. Importantly, we propose computing the weights in terms of patches rather than individual pixels. 

    \item We introduce upsampling and downsampling layers designed for a general sampling factor, leveraging the geometric information from the PAN data.
    
    \item We also present a new post-processing module for image enhancement, which uses the proposed multi-head attention residual architecture to perform nonlocal filtering. Although this module is used in the context of image fusion, it can be readily adapted to other image processing tasks.

    \item A comprehensive experimental section is conducted, testing the generalization capabilities of our method and numerous state-of-the-art fusion techniques. We use data from PRISMA, Quickbird, and WorlView2 satellites, which each have different sensor configurations.
\end{itemize}

Table \ref{tab:notation} contains a list of abbreviations and notations used throughout this work.

The rest of the paper is organized as follows. In Section \ref{sec:related}, we review the state of the art in satellite image fusion. Section \ref{sec:proposed} introduces the proposed model-based deep unfolded method for pansharpening and hypershaprening, while Section \ref{sec:implementation} provides the implementation details. An extensive performance comparison across PRISMA, Quickbird and WorldView-2 datasets is presented in Section \ref{sec:experimentation}. Section \ref{sec:ablation} conducts an ablation study that highlights the selected network configurations and assesses the robustness of our method to variations in sampling factor and noise. Finally, conclusions are drawn in Section \ref{sec:conclusion}.

\section{Related Work}\label{sec:related}

A wide variety of methods have been proposed in the literature for pansharpening and hypersharpening. These methods can be roughly categorized into three groups: classic methods, pure learning-based methods, and model-based deep unfolding methods.

\subsection{Classic methods}
Within classic methods, we find component substitution algorithms, multiresolution analysis approaches, and variational optimization techniques. Component substitution methods aim to replace a specific component of the MS/HS image with the corresponding component extracted from the PAN image. For this purpose, several approaches have been proposed, such as Principal Component Analysis (PCA) \cite{kwarteng1989extracting, chavez1991comparison}, Brovey \cite{gillespie1987color},
Intensity Hue Saturation (IHS) \cite{carper1990use, chavez1991comparison}, and Gram-Schmidt (GS) \cite{aiazzi2007improving}.

Multiresolution analysis approaches inject the spatial details acquired through a multiscale decomposition from the PAN image into the MS/HS data. The multiscale decomposition can be carried out by different techniques. Laplacian pyramid \cite{aiazzi2002context, aiazzi2006mtf,lee2009fast}, contourlet transform \cite{shah2008efficient}, curvelet transform \cite{nencini2007remote}, discrete wavelet transform \cite{yocky1995image, nunez1999multiresolution,ranchin2000fusion, otazu2005introduction}, and high-pass modulation \cite{de1998fusion,khan2009pansharpening} are mainly used. 

Hybrid methods combining the component substitution and multiresolution  mechanisms have also been proposed \cite{li2005color, luo2008fusion, cheng2015remote}.

Finally, variational optimization methods address the fusion problem by minimizing an energy functional derived from observation models and prior knowledge about the expected solution. The minimization is achieved using diverse techniques, such as proximal gradient \cite{chambolle2016introduction}, Alternating Direction Method of Multipliers (ADMM) \cite{boyd2011distributed,parikh2014proximal}, and primal-dual algorithms \cite{chambolle2011first}, among others. In this setting, several works can be found for pansharpening and hypershapening based on different regularization and data terms \cite{ballester2006variational,palsson2012new,duran2014nonlocal,he2012pansharpening}.

\subsection{Pure deep learning-based methods}
Over the past decade, an increasing number of deep learning-based image fusion methods have been proposed in the literature, showing promising results. These approaches can be categorized based on their architectures. In this context, we find generative adversarial networks \cite{liu2020psgan}, but primarily residual connections \cite{Yang_2017_ICCV, yuan2018multiscale, he2019pansharpening, deng2020detail, cai2020super} and attention modules \cite{bandara2022hypertransformer, zhou2022panformer}.

Several approaches for image fusion base their architectures on residual connections. PanNet \cite{Yang_2017_ICCV} performs the fusion by adding the output of a convolutional neural network to the bicubic interpolation of the MS image. Specifically, the network receives the PAN and MS images, computes a high-pass filter from them, and then passes the results through a residual network. MSDCNN \cite{yuan2018multiscale} extracts multiscale features using residual connections to preserve both the coarse structure and texture details. DiCNN \cite{he2019pansharpening} injects details extracted from a concatenation of the PAN and upsampled MS/HS images using a block composed of convolutional layers and ReLU activation functions. FusionNet \cite{deng2020detail} estimates the geometric details from the difference between the PAN and upsampled MS/HS data using residual blocks and then injects them into the upsampled MS/HS image. SRPPNN \cite{cai2020super} adapts a residual architecture for super-resolution to pansharpening by incorporating the PAN image at different scales and injecting its high frequencies into the processed upsampled MS/HS data.

Inspired by nonlocal networks \cite{wang2018non}, attention modules \cite{dosovitskiy2020image} and leverage image self-similarities are a key component of transformers blocks. In this context, HSIT \cite{bandara2022hypertransformer} introduces a transformer-based mechanism for pansharpening that extracts features from HS and PAN images to generate queries and keys. The feature extraction process is carried out using a VGG-like architecture \cite{simonyan2014vgg}. Subsequently, a sequence of transformer blocks is applied at different scales, and a fusion module is added at the end. On the other hand, PanFormer \cite{zhou2022panformer} proposes a dual-path encoder for modality-specific feature extraction applied to MS and PAN images within a self-attention block. Each encoder is built with a stack of these blocks to produce intermediate features from the input. Then, a cross-modality module merges the spectral and spatial features. Finally, the pansharpened image is obtained by enhancing the output from the previous module.

Finally, AWFLN \cite{lu2023awfln} combines residual connections with self-attention mechanisms. The algorithm estimates the fused image by adding missing spatial details to the upsampled MS/HS data. To achieve this, the authors combine a multiscale convolution block with adaptive convolution. These convolutions are adapted to the data by computing the weights of the kernel using spatial and spectral attention mechanisms.

\subsection{Model-based deep unfolding methods}

While the use of observation models makes variational methods robust to distortions, their performance is constrained by rigid priors. Conversely, purely data-driven learning approaches can adeptly learn natural priors but are less flexible and interpretable than model-based techniques. Deep unfolding combines the strengths of both paradigms. The general idea involves unrolling the steps, commonly referred to as {\it stages}, of the optimization scheme derived from the minimization of an energy functional into a deep learning framework, resulting in efficient and interpretable architectures. Neural networks typically replace specific operations within the scheme. The learnable architectures are either pretrained or scheduled for training. Recently, several works for image fusion have been proposed in this setting \cite{xie2020mhf,xu2021deep,yan2022mmnet,mifdal2023deep,zhang2023spatial,li2023local,mai2024deep}.

For clarity, we recall some of the basic notations introduced in Table \ref{tab:notation}. We denote the MS/HS image as $\HS \in \R^{C\times h\times w}$, the PAN image as $\PAN\in \R^{H\times W}$, and the desired fused image as $\U \in \R^{C\times H \times W}$. Here, $C$ is the number of spectral bands, $w$ and $h$ represent the width and height of the low-resolution data, while $W$ and $H$ denote those of the high-resolution image. These notations are compiled in the Table \ref{tab:notation}.

The classic observation model \cite{ballester2006variational} that relates the fused image to the low-resolution MS/HS data is given by
\begin{equation}\label{eq:DBu}
\HS = \DB (\U) + \eta,    
\end{equation}
where $\text{B}$ is a low-pass filter, $\text{D}$ is the $s$-decimation operator, $s$ is the sampling factor between the low and high-resolution domains, and $\eta$ is the noise realization. Furthermore, it is commonly assumed that the PAN image is obtained as a linear combination of the MS/HS spectral components \cite{ballester2006variational}, that is,
\begin{equation}\label{eq:p_linear}
    \PAN = \sum_{i=1}^C \alpha_i \U_i,
\end{equation}
where $\U_i$ denotes the $i$-th channel of the desired image and $\{\alpha_i\}$ are mixing coefficients that depend on the spectral sensitivity of the sensors.

MHF-net \cite{xie2020mhf} formulates an energy fidelity term based on \eqref{eq:DBu} and \eqref{eq:p_linear} and unmixing theory \cite{li2010new}. By combining both models, they derive a minimization problem in terms of the low-rankness representation of the high-resolution HS image. This is achieved by unfolding the minimization scheme obtained through the proximal gradient algorithm \cite{parikh2014proximal}. The associated operators are replaced with convolutional neural networks. Additionally, the authors introduce a novel architecture based on residual networks, where upsampling operators are guided by the PAN image. 

Xu et al.~\cite{xu2021deep} introduced GPPNN, which is also based on the observation models \eqref{eq:DBu} and \eqref{eq:p_linear}. They define an energy functional for each data term with their own regularization. The image fusion is performed by solving a two-level minimization problem using the proximal gradient algorithm \cite{parikh2014proximal}. The steps of the iterative schemes are computed sequentially, with the output of one step serving as the input for the next one. The proximal and other operators are replaced with convolutional layers.

MMNet \cite{yan2022memory} proposes an energy formulation with a fidelity-term based on \eqref{eq:DBu} and two regularization terms: one local and one nonlocal. The authors introduce an auxiliary variable and solve the resulting minimization problems using the proximal gradient algorithm \cite{parikh2014proximal}. All operators are replaced with convolutional neural networks. The architectures that replace the regularization terms are similar, with the distinction that the nonlocal one applies a nonlocal filter to the input variable, guided by the PAN image. Finally, a memory mechanism is introduced to retain information from previous stages.

The HyFPan model of Mifdal et al.~\cite{mifdal2023deep} introduces an energy functional comprising three terms: the classic observation model \eqref{eq:DBu}, a radiometric constraint pioneered by Duran et al.~\cite{duran2017survey} to recover the geometry of the scene, and a regularization term. The radiometric constraint ensures that the high frequencies of each spectral band in the fused image align with those of the PAN data. Additionally, HyFPan leverages the low-rank representation of the high-resolution MS/HS image as proposed in \cite{xu2021deep}. For minimization, the authors employ the proximal gradient algorithm \cite{parikh2014proximal} and substitute all involved operators with convolutional neural networks.

LGTEUN \cite{li2023local} integrates transformer layers with unfolding mechanisms to address the minimization of an energy derived from \eqref{eq:DBu} and \eqref{eq:p_linear}. The proposed method unfolds the proximal gradient iterative scheme and introduces a network architecture that combines a lightweight CNN data module with a transformer module. To effectively capture both local and global dependencies, the authors incorporate two branches within the transformer module: a local branch and a global branch. The local branch employs multi-head self-attention within spatial windows to focus on local features, while the global branch extracts global contextual features in the frequency domain, leveraging the properties of the Fourier transform.

Mai et al.~\cite{mai2024deep} introduce the UTeRM model. Similar to MHF-net, the authors propose an energy formulation for pansharpening through low-rank representation. As a novelty, they evaluate three different constraints for detail injection inspired by CS, MRA, and CNN-based formulations. 

In contrast to the previous deep unfolding methods, S2DBPN \cite{zhang2023spatial} is inspired by the back-projection method \cite{irani1993motion, dai2007bilateral}. This approach is based on the assumption that a recovered image can be obtained by minimizing the reconstruction error between the target image and the observed data. Based on \eqref{eq:DBu} and \eqref{eq:p_linear}, the authors design two back-projection schemes that are unfolded in parallel. At each stage, the results of both back projections are averaged and used as input for the subsequent stage. Ultimately, the outputs of all stages are combined to produce the final fused image.

\section{Proposed Method}\label{sec:proposed}

From now on, $\PAN$ will denote the PAN image replicated across all spectral bands, that is, $\PAN \in\R^{C \times H \times W}$ such that $\PAN_j = \PAN_k$ for all $j,k \in \{1,\cdots, C\}$. 

While the observation model \eqref{eq:DBu} is widely used, several alternatives to \eqref{eq:p_linear} have been proposed in the literature to inject the high frequencies from the PAN image to the fused result. We adopt the constraint introduced in \cite{duran2017survey}, which assumes that the high frequencies of each spectral band of the fused image are proportional to those of the PAN. In this context, the high frequencies of $\PAN$ are estimated as $\PAN-\widehat{\PAN}$, where $\widehat{\PAN}\in\R^{C\times H\times W}$ represents the low frequencies of the PAN, obtained by applying bicubic interpolation to the downsampling of $\PAN$. Similarly, the high frequencies of the fused image are obtained as $\U-\widehat{\HS}$, where $\widehat{\HS}\in\R^{C\times H\times W}$ is the result of upsampling $\HS$ by bicubic interpolation. Therefore, we impose the following constraint:
\begin{equation}\label{eq:radiometric}
    \U_k -\widehat{\HS}_k = (\widehat{\HS}_k\oslash\widehat{\PAN}_k)\odot(\PAN_k-\widehat{\PAN}_k), \:\: \forall k \in \{ 1,\cdots, C\},
\end{equation}
where $\odot$ and $\oslash$ denote the element-wise product and quotient, respectively, and $\widehat{\HS}_k\oslash \widehat{\PAN}_k$ is a modulation coefficient that accounts for the energy of each spectral band.

\subsection{Variational formulation}

In order to estimate the fused image from \eqref{eq:DBu} and \eqref{eq:radiometric}, we propose the following variational model:
\begin{equation}\label{eq:energy_def}
   \min_{\U} \frac{\lambda}{2}\| \DB(\U)-\HS\|^2_{2} + \beta  \|\widehat{\PAN}\odot \U -   \PAN\odot \widehat{\HS}\|_{1} + \mu \mathcal{R(\U)},
\end{equation}
where $\lambda, \beta,\mu>0$ are trade-off parameters and $\mathcal{R}(\U)$ is an arbitrary convex regularization term that promotes smoothness of the solution. Note that we have rewritten the constraint \eqref{eq:radiometric} as
\begin{equation}\label{eq:energy_rad}
\widehat{\PAN}\odot \U = \PAN\odot\, \widehat{\HS}.
\end{equation}
The choice of the squared $L^2$ norm in \eqref{eq:DBu} is based on the assumption that the noise follows a Gaussian distribution. Conversely, the data involved in \eqref{eq:energy_rad} exhibits noise correlation between spectral bands due to downsampling and bicubic operations. In such cases, the $L^1$ norm offers greater robustness.

Since the minimization problem \eqref{eq:energy_def} is convex but non-smooth, we use the first-order primal-dual algorithm proposed by Chambolle and Pock \cite{chambolle2011first} to efficiently find a global optimal solution\footnote{We refer the reader to \cite{chambolle2016introduction} for all details on convex analysis omitted in this subsection.}. For that purpose, we reformulate \eqref{eq:energy_def} into a saddle-point problem using the dual variables $\V\in \R^{C\times H \times W}$ and $\TT\in \R^{C\times h \times w}$ as follows: 
\begin{equation}\label{eq:saddle-point}
\begin{aligned}
\min_{\U} \max_{\V,\TT}  &\langle \U , \BU ({\TT}) + \widehat{\PAN}\odot{\V} \rangle -\langle \HS, {\TT}\rangle -\frac{1}{2\lambda}\|{\TT}\|^2_{2} \\
&-\langle \PAN \odot \widehat{\HS} ,{\V}\rangle- \delta^{*}_{\beta}({\V}) +\mu \mathcal{R}(\U),
\end{aligned}
\end{equation}
where we have assumed that $\text{B}^t=\text{B}$. The indicator function $\delta_{\beta}^{*}:  \R^{C\times H \times W}\rightarrow \R$ is defined as
$$
\delta_{\beta}^{*}({\V}) = \begin{cases}
    0 & \text{ if } \| \V \|_{\infty} \leq \beta, \\
    +\infty & \text{ otherwise}.
\end{cases}
$$

The primal-dual Chambolle-Pock algorithm uses the proximity operator, which generalizes the projection onto convex sets. For a proper convex function $f$, it is defined as
\begin{equation}\label{eq:prox}
\begin{aligned}
\prox_{\tau f}(x)&=\argmin_y \left\{ f(y)+\dfrac{1}{2\tau}\|x-y\|^2_2\right\} \\
&= (\text{Id} + \tau \partial f)^{-1}(x),
\end{aligned}
\end{equation}
where $\tau>0$ is the step-size and $(Id + \tau \partial f)^{-1}$ is called the resolvent operator \cite{chambolle2016introduction}. The algorithm consists in an ascent step in the dual variables and a descent step in the primal variable, followed by over-relaxation to ensure convergence. The resulting primal-dual optimization scheme for \eqref{eq:saddle-point} is 
\begin{equation}\label{eq:iter_scheme}
\begin{cases}
{\TT}^{n+1} = & \frac{1}{1+\tau_d/\lambda}  ({\TT}^n +\tau_d \DB(\overline{\U}^n )- \tau_d\HS),\\ 
{\V}^{n+1} = &   \dfrac{{\V}^n +\tau_d \widehat{\PAN}\odot \overline{\U}^n - \tau_d \PAN\odot \widehat{\HS}}{\max({\V^n}+\tau_d \widehat{\PAN}\odot\overline{\U}^n-\tau_d \PAN \odot\widehat{\HS}, \beta)}, \\ 
\U^{n+1} = & \prox_{\tau_p\mu \mathcal{R}} ({\U}^n -\tau_p \BU( {\TT^{n+1}}) -\tau_p \widehat{\PAN} \odot {\V^{n+1}}), \\ 
\overline{\U}^{n+1} = & 2 \U^{n+1} -\U^n,
\end{cases}
\end{equation}
where $\tau_p>0$ and $\tau_d>0$ are the primal and dual step-size parameters, respectively.

\subsection{Unfolded formulation}

We propose to unfold \eqref{eq:iter_scheme} and replace the involved operators with learning-based networks. Hereafter, we refer to each step $n$ in the iterative scheme as a {\it stage}. We replace $\DB$ with $\Down^n$ and $\BU$ with $\Up^n$, which are detailed in Subsection \ref{subsec:upsampling}. Furthermore, the proximity operator in \eqref{eq:iter_scheme} related to the regularization term is replaced by the multi-head attention residual network proposed in Subsection \ref{sec:marnet} and denoted by $\MARNet^n$. This framework avoids the need to handcraft the prior $R$ while exploiting self-similarity in natural images within nonlocal operators \cite{gilboa2009nonlocal,buades2005non,wang2018non}. Accordingly, the unfolded version of the primal-dual scheme \eqref{eq:iter_scheme} results:

\begin{equation}\label{eq:iter_scheme_unfolded}
\begin{cases}
{\TT}^{n+1} = & \frac{1}{1+\tau_d/\lambda}  ({\TT}^n +\tau_d \Down^{n+1}(\overline{\U}^n )- \tau_d\HS),\\ 
{\V}^{n+1} = &   \dfrac{{\V}^n +\tau_d \widehat{\PAN}\odot \overline{\U}^n - \tau_d \PAN\odot \widehat{\HS}}{\max({\V^n}+\tau_d \widehat{\PAN}\odot\overline{\U}^n-\tau_d \PAN \odot\widehat{\HS}, \beta)}, \\ 
\U^{n+1} = & \MARNet^{n+1} ({\U}^n -\tau_p \Up^{n+1}( {\TT^{n+1}}) -\tau_p \widehat{\PAN} \odot {\V^{n+1}}), \\ 
\overline{\U}^{n+1} = & 2 \U^{n+1} -\U^n.
\end{cases}
\end{equation}
 
The networks $\Down^n$, $\Up^n$ and $\MARNet^n$ do not share weights between the different stages. However, the hyperparameters $\lambda$, $\beta$, $\mu$, $\tau_p$ and $\tau_d$ are shared across all stages and  learned during the training phase. 

\begin{figure}[t]
    \centering
    \includegraphics[ trim=0.5cm 0cm 3cm 0cm, clip, width=\linewidth]{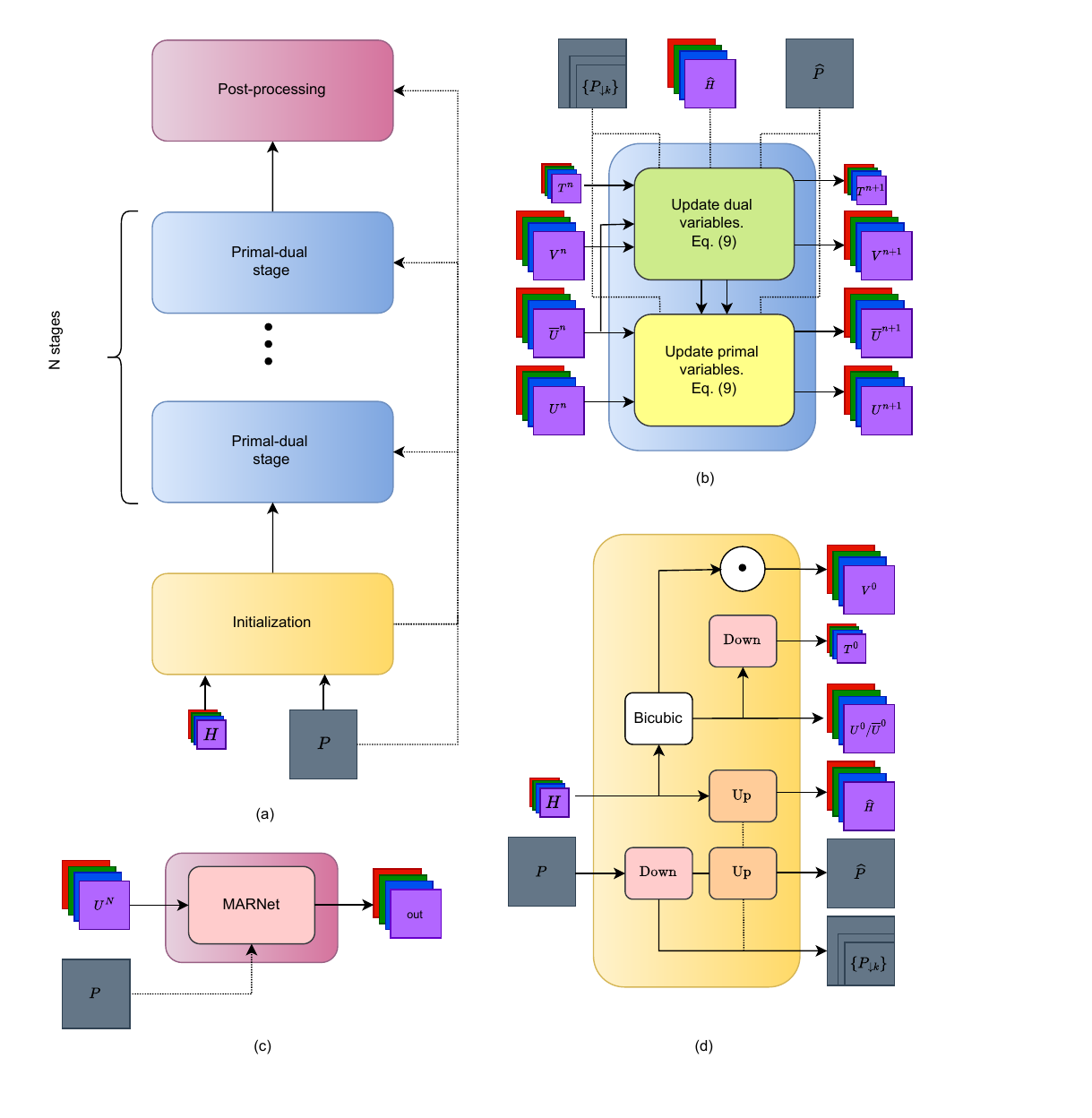}
    \caption{(a) Overall architecture of the proposed method. (b) A single primal-dual stage. (c) Post-processing module. (d) Initialization module.}
    \label{fig:overall}
\end{figure}

Figure \ref{fig:overall}a illustrates the overall architecture. Figure \ref{fig:overall}b shows one single stage. Figure \ref{fig:overall}c displays the post-processing that occurs after the last stage. Figure \ref{fig:overall}d depicts the initialization module that learns the auxiliar data $\widehat{\HS}$ and $\widehat{\PAN}$ involved in the radiometric constraint \eqref{eq:energy_rad}. In the subsequent subsections, we provide more details about all networks and modules.
 
\subsection{Upsampling and downsampling operators}\label{subsec:upsampling}

Inspired by \cite{xie2020mhf}, for a given sampling factor $s\in\mathbb{Z}^+$, we consider its prime decomposition $s=q_1 \cdot q_2\cdots q_M$ such that $q_1\leq q_2\leq \cdots\leq q_M$. We proceed in $M$ steps, iterating over the primes in descending order for downsampling and in ascending order for upsampling.

For downsampling, we use the decimation $D_q$ with a sampling rate of $q$ and a 2D convolution $\mathcal{C}$. Therefore, one step is obtained by $\mathcal{D}_q =D_q\circ \mathcal{C}$, and the downsampling operator is  defined as the composition
\begin{equation*} 
\Down = \mathcal{D}_{q_1} \circ  \mathcal{D}_{q_2}\circ \cdots \circ\mathcal{D}_{q_M}.
\end{equation*}

Following the classic adjoint relationship between upsampling and downsampling, we define the upsampling operator as
\begin{equation*}
    \Up = \mathcal{C}\circ \mathcal{D}^{\,t}_{q_M} \circ\cdots\circ \mathcal{D}^{\,t}_{q_2}\circ \mathcal{D}^{\,t}_{q_1},
\end{equation*}
where a final convolution $\mathcal{C}$ is added. We propose leveraging the spatial information from the PAN data according to the corresponding resolution level. For this purpose, we consider the sequence $\{\PAN_{\downarrow k}\}_{k=0}^{M-1}$, where $\PAN_{\downarrow 0}=\PAN$ and
\begin{equation}\label{eq:Pseq}
\PAN_{\downarrow k} = \mathcal{D}_{q_{M-k+1}}\left(\PAN_{\downarrow k-1}\right)
\end{equation}
for $k\geq 1$. Therefore, one upsampling step is defined as $\mathcal{D}^{\,t}_{q_i}=\mathcal{G}_{P_{\downarrow (M-i)}}\circ\mathcal{C}_{q_i}^{\,t}$, where $\mathcal{C}^t_{q_i}$ is a transposed 2D convolution with a stride of $q_i$, and $\mathcal{G}_{P_{\downarrow (M-i)}}$ is a geometry injection module. This module concatenates its input with the corresponding $\PAN_{\downarrow (M-i)}$, which contains an accurate description of the geometry at the given resolution, and applies three layers composed of a convolution and batch normalization. 

\begin{figure}[t!]
    \centering
    \includegraphics[ trim=1.5cm 0.5cm 1cm 0.5cm, clip, width=\linewidth]{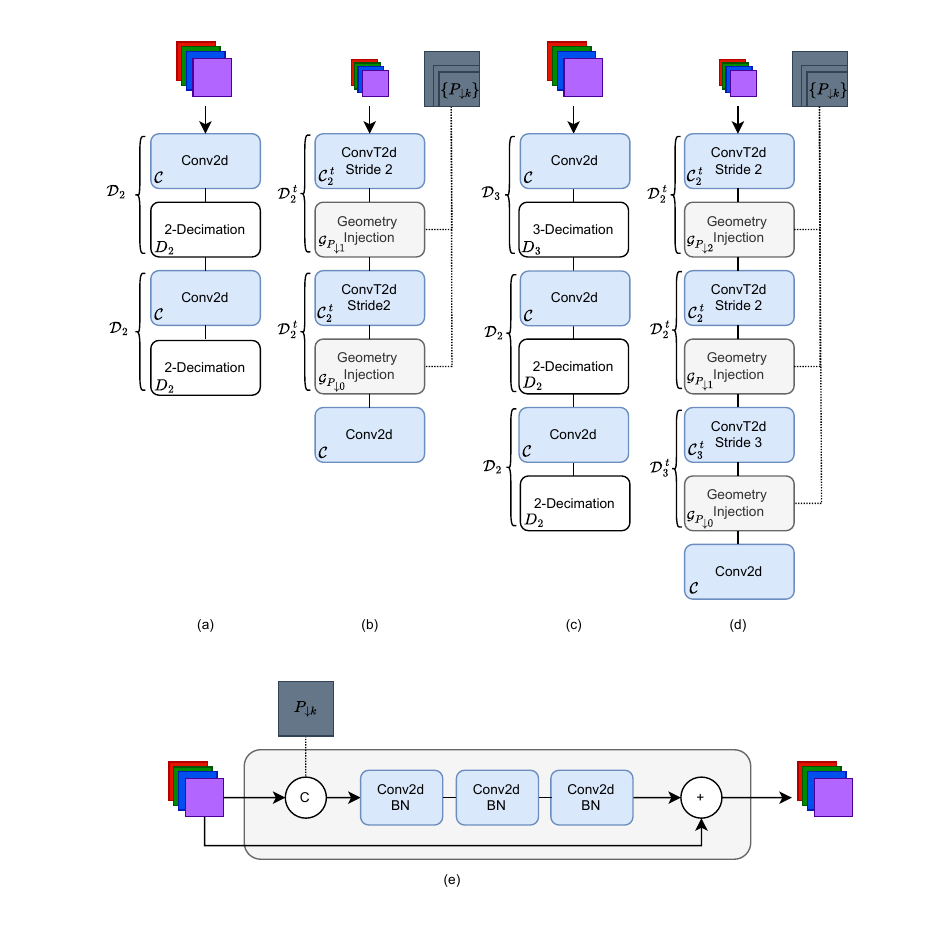}
    \caption{Diagrams (a) and (b) display the Down and Up operators for sampling 4, while (c) and (d) show those for sampling 12. Diagram (e) depicts the architecture of the proposed geometry injection layer.}
    \label{fig:upsampling}
\end{figure}

Figure \ref{fig:upsampling} presents the architecture of the $\Down$ and $\Up$ operators for sampling rates of 4 and 12.

\subsection{Multi-head Attention Residual Network (MARNet)}\label{sec:marnet}

\begin{figure*}[t]
    \centering
\begin{tabular}{c}
\includegraphics[ trim=0cm 0cm 0cm 0cm, clip, width=0.90\linewidth]{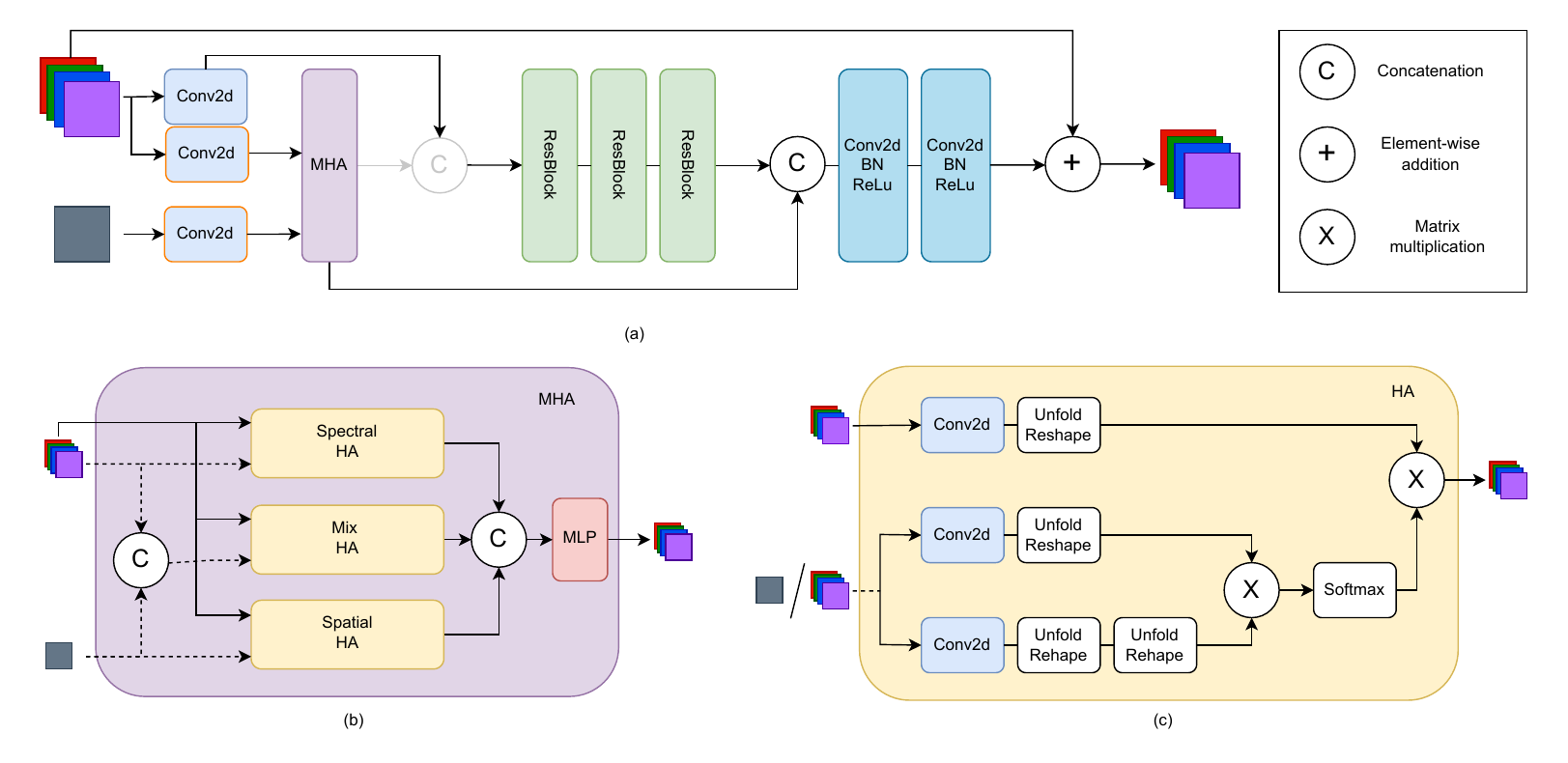}\\
\end{tabular}
    \caption{(a) Proposed MARNet. (b) Multi-Head Attention (MHA) module, which is composed of three Head Attention (HA) layers. (c) A single HA layer.}
    \label{fig:marnet}
\end{figure*}

If in \eqref{eq:prox} we assume that $\tau$ is sufficiently small and that $f$ is differentiable, then the proximity operator can be approximated by \cite{chambolle2016introduction}
\begin{equation}\label{eq:prox2}
    \prox_{\tau f}(x) \approx x - \tau \nabla f(x).
\end{equation}
Accordingly, proximity operators are typically close to the identity and, thus, replaced by residual networks when unfolding. 

In view of \eqref{eq:prox2}, we present a novel architecture that integrates a residual network with a multi-head attention module, designed to exploit image self-similarities. We refer to our proposal as Multi-Head Attention Residual Network (MARNet), and its architecture is illustrated in Figure \ref{fig:marnet}.

As depicted in the figure, we first apply a multi-head attention module (MHA) to input and PAN image features, followed by three residual blocks (ResBlocks). Each of these blocks consists of two convolutional layers followed with a residual connection. The attention and residual features are then processed through two convolution layers, whose output is added to the input data.

The multi-head attention module comprises multiple head attentions, each of which approximates the nonlocal means filter \cite{buades2005non}:
\begin{equation*}
\text{NL}(g)_i = \sum_{j} \omega_{i,j} g_{j},
\end{equation*}
where $\omega_{i,j}$ are the weights that measure the similarity between $g_i$ and $g_j$. The weights are commonly computed in terms of the Euclidean distance between patches $Q_i$ and $Q_j$ centered at pixels $i$ and $j$, respectively:
$$
\omega_{i,j} = \dfrac{1}{\Gamma_i} \exp\left(- \frac{\left\| Q_i-Q_j\right\|^2}{h^2_{\text{sim}}}\right),
$$
where $\Gamma_i$ is a normalization factor and $h_{\text{sim}}>0$ is a filtering parameter that measures how fast the weights decay with increasing dissimilarity between patches. In the nonlocal networks introduced by Wang et al.~\cite{wang2018non}, the Euclidean distance between patches is replaced with the scalar product between pixels, and the filtering parameters are learned through convolutions. Combining these ideas, we propose to compute the similarity weights as
\begin{equation}\label{eq:nlweights}
\omega_{i,j} = \dfrac{1}{\Gamma_i} \exp\left(\theta(Q_i)^t \phi(Q_j) \right).
\end{equation}
where $\theta$ and $\phi$ are learnable convolutions. Furthermore, we restrict the non-zero weights to a window of size \(r\) centered at each pixel, that is, \(\omega_{i,j} = 0\) for $j$ such that \(\|i-j\|_{\infty} > r\).

The proposed multi-head attention mechanism comprises three attention heads arranged in parallel. Each head computes the weights \eqref{eq:nlweights} on a different auxiliary image: the input, the PAN image, and a concatenation of both. This design enables the network to process spectral, spatial, and combined information, thereby enhancing its ability to focus on relevant features during the unfolding process. Subsequently, a multi-layer perceptron \cite{cybenko1989approximation} is used to integrate the information from all heads. Figure \ref{fig:marnet}b illustrates the proposed multi-head attention module, while Figure \ref{fig:marnet}c depicts the architecture of a single head.

\subsection{Initialization and post-processing}\label{sec:init}

In the initialization stage, the operators $\Down^0$ and $\Up^0$ are used to generate the low frequencies of $\HS$ and $\PAN$, denoted as $\widehat{\HS}$ and $\widehat{\PAN}$, respectively. Furthermore, the sequence $\{\PAN_{\downarrow k}\}_{k=0}^{M-1}$ is computed according to \eqref{eq:Pseq} and the primal and dual variables in \eqref{eq:iter_scheme_unfolded} are initialized. Ultimately, the initializations are
\begin{equation*}
\begin{split}
    \widehat{\PAN} &= \Up^0(\Down^0(\PAN)),\\
    \widehat{\HS} &= \Up^0(\HS), \\
    \overline{\U}^0=\U^0 &=\Bicubic(\HS),\\
    {\TT}^0   &= \Down^0(\U^0),\\
    {\V}^0 &= \U^0 \odot \widehat{\PAN}.
\end{split}
\end{equation*}

As a post-processing step, after completing a fixed number of $N$ stages, we apply a MARNet layer to $\U^N$ to obtain the final fused image $\U_{out}$.

\section{Implementation Details}\label{sec:implementation}

The proposed deep unfolded network is trained during 1000 epochs using the following loss function:
\begin{equation}\label{eq:lossFinal}
\begin{split}
    \text{Loss}(\U_{out}, \{\U^i\}_{i=1}^N, \text{GT}) &= \text{L1}(\U_{out}, \text{GT}) \\
    &+ \frac{\alpha}{N} \sum_{i=1}^N\text{MSE}(\U^{i}, \text{GT}),
\end{split}
\end{equation}
where $\text{GT}$ represents the ground-truth image, $\text{MSE}$ denotes the mean squared error, and $\alpha$ is fixed to $0.1$. Note that we combine the L1 loss between the final fused image $\U_{out}$ and the ground truth with the MSE at each stage. We employ Adam optimizer {\cite{boyd2011distributed,parikh2014proximal}}, which is known for its efficiency in handling sparse gradients and adapting learning rates individually for each parameter. This makes it particularly well-suited for complex and high-dimensional datasets. The initial learning rate was set to $5\cdot 10^{-4}$.  We also set the number of primal-dual stages to $N=4$. 

Subsequently, we fine-tuned the model by retraining for an additional 100 epochs. During this phase, the weights of the initialization module and the primal-dual stages were kept fixed, and only the parameters involved in the post-processing were trained. In this case, we use the L1 loss between the output and the ground truth.

During the training phase, at the end of each epoch, we evaluate the model on the validation set. If the PSNR outperforms the previous best value, the weights are updated accordingly. Consequently, the final weights correspond to the parameters that achieved the highest PSNR on the validation set.

\section{Experimental Results}\label{sec:experimentation}

\begin{figure*}[ht]
    \centering
    \begin{tabular}{c@{\hskip 0.2em}c}
         \includegraphics[width=0.45\linewidth]{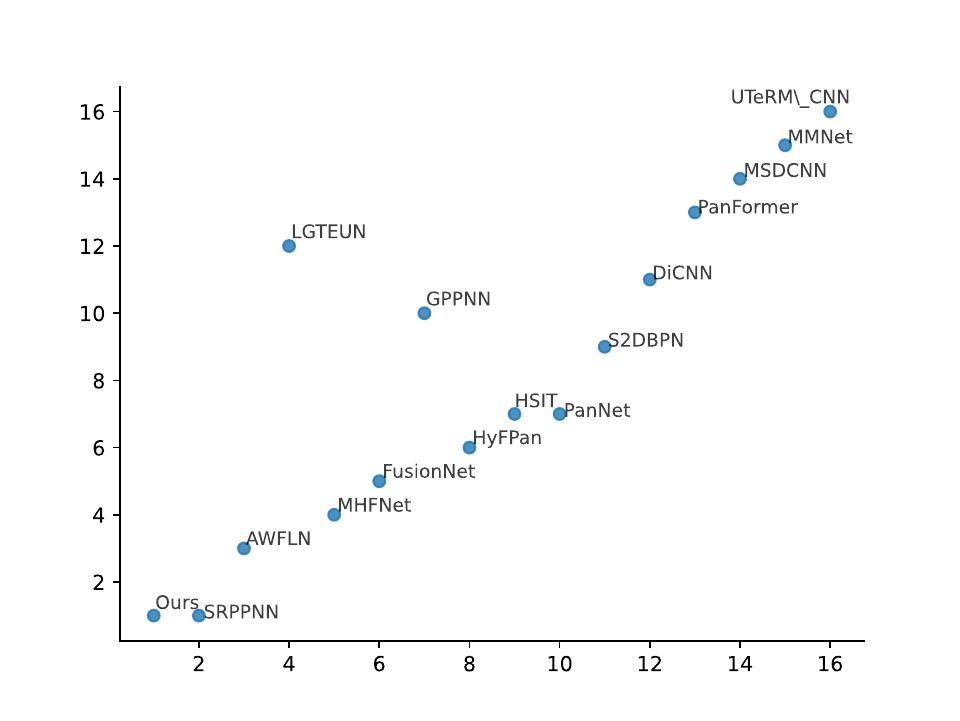} & 
         \includegraphics[width=0.45\linewidth]{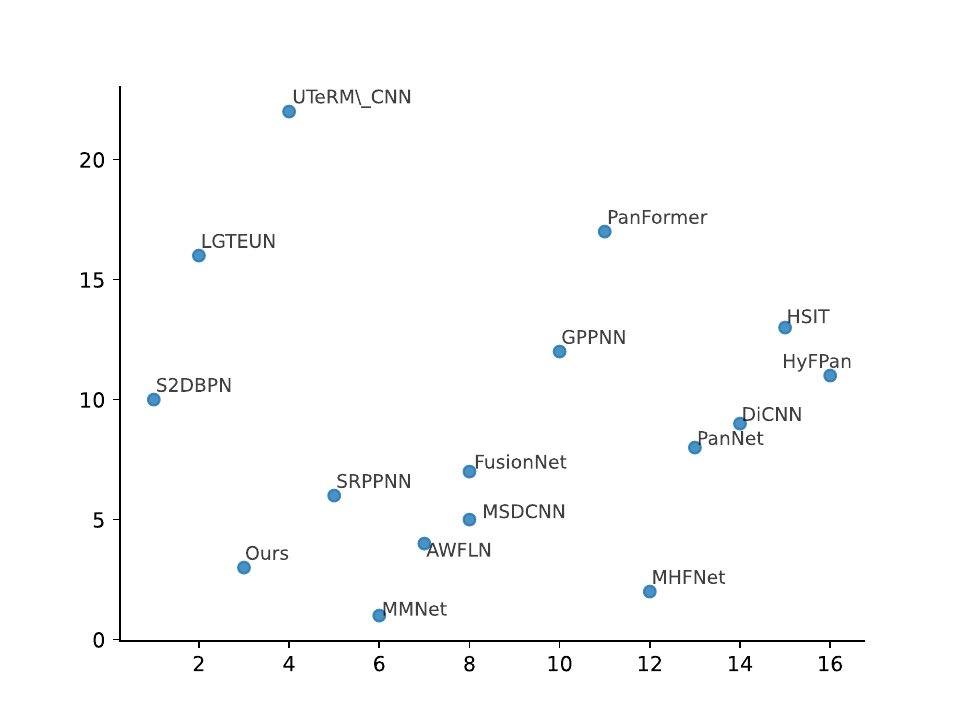} \\
         (a) Prisma & (b) QuickBird  \\
         \includegraphics[width=0.45\linewidth]{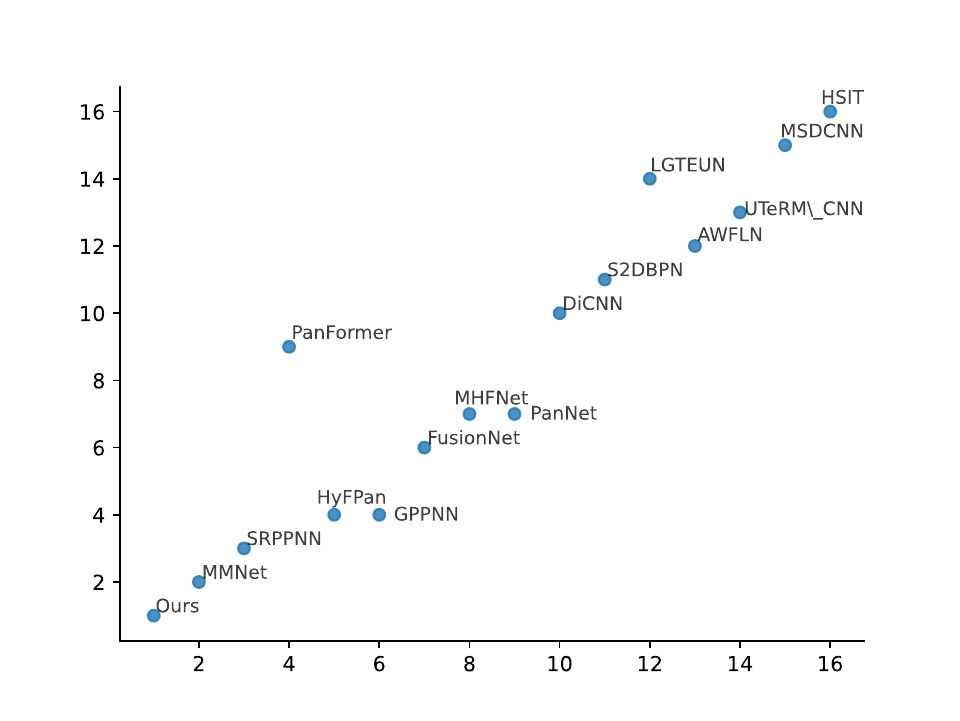} &
         \includegraphics[width=0.45\linewidth]{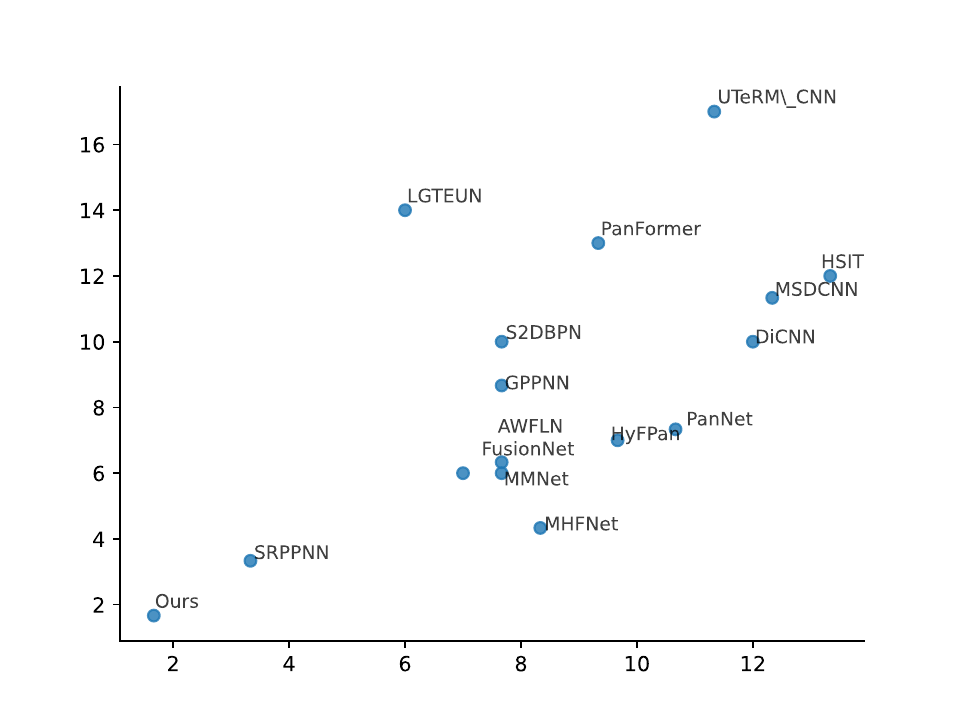} \\
        (c) WorldView2 & (d) Mean\\
        
    \end{tabular}
    \caption{Each graphic compares the rank positions based on the PSNR between the validation (horitzontal axis) and the testing (vertical axis) sets. Accordingly, a method positions lower and further to the left indicates better performance. In (d), the mean rank across the three datasets is displayed. It is observed that the proposed fusion method performs the best in all cases.}
    \label{fig:rank-metrics}
\end{figure*}

In this section, we evaluate the performance of the proposed pansharpening method by analyzing numerical and visual results. To ensure a comprehensive assessment, we conduct experiments on a large variety of data, testing the generalization capabilities of competing techniques. Specifically, we use three different datasets, one for each of the Earth observation satellites Prisma, QuickBird, and WorldView2. 

We compare our approach with various state-of-the-art methods: bicubic interpolation; the classic methods Brovey \cite{gillespie1987color}, IHS \cite{chavez1991comparison}, GSA \cite{aiazzi2007improving}, Wavelet \cite{king2001wavelet}, PCA \cite{kwarteng1989extracting}, and CNMF \cite{yokoya2011coupled}; the pure deep learning-based approaches PanNet \cite{Yang_2017_ICCV}, MSDCNN \cite{yuan2018multiscale}, DiCNN \cite{he2019pansharpening}, FusionNet \cite{deng2020detail}, SRPPNN \cite{cai2020super}, HSIT \cite{bandara2022hypertransformer}, Panformer \cite{zhou2022panformer}, and AWFLN \cite{lu2023awfln}; and the model-based deep unfolding methods MHFNet \cite{xie2020mhf}, GPPNN \cite{xu2021deep}, MMNet \cite{yan2022mmnet}, HyFPan \cite{mifdal2023deep}, S2DBPN \cite{zhang2023spatial}, LGTEUN \cite{li2023local}, and UTeRM\_CNN \cite{mai2024deep}. The codes for the classic methods were obtained from the Py\_pansharpening toolbox\footnote{\url{https://github.com/codegaj/py_pansharpening/tree/master}}, while those for PanNet, MSDCNN, DiCNN, and FusionNet were sourced from the DLPan-toolbox\footnote{\url{https://github.com/liangjiandeng/DLPan-Toolbox/tree/main}}. All other codes were downloaded from the websites of the corresponding authors.

For quantitative comparison, we use the following metrics: ERGAS (Erreur Relative Globale Adimentionelle de Synth\`ese) \cite{rabbani1991digital, ranchin2000fusion}, which measures the global spatial quality, PSNR (Peak Signal to Noise Ratio) \cite{rabbani1991digital}, which assesses the spatial reconstruction quality with respect to noise, SSIM (Structural Similarity Index Measure) \cite{wang2002universal}, which evaluates the overall quality of the fused image, SAM (Spectral Angle Mapper) \cite{alparone2004global}, which measures the spectral reconstruction quality, and Q$2^n$ \cite{garzelli2009hypercomplex}, which evaluates the loss of correlation, luminance and contrast distortion. 

The state-of-the-art techniques based on pure deep learning and model-based unfolding were trained using the loss functions and training hyperparameters specified in the respective articles. The weights used for the evaluation correspond to the parameters that achieved the best performance in terms of PSNR on the validation set.

\subsection{Experiments on Prisma}

In 2019, the Italian Space Agency initiated the PRISMA mission, dedicated to providing public data for specific areas, with significant potential for fusion and resolution enhancement. The HS data includes 66 channels corresponding to the Visible and Near Infrared (VNIR) spectrum, covering wavelengths between 400 and 1010 nm, and 173 channels corresponding to the Short Wavelength Infrared (SWIR) spectrum, covering wavelengths between 920 and 2505 nm. All of these spectral bands have a resolution of 30 m per pixel. The PAN image consists of a single channel with a spatial resolution of 5 m, which also covers the same spectrum as VNIR.

Using the PRISMA mission portal\footnote{\url{https://www.asi.it/en/earth-science/prisma/}}, we collected and downloaded 14 large-scale images. Their original size was $1000\times 1000\times 239$ for the HS data and $6000\times 6000$ for the PAN image. The first 66 bands correspond to the bandwidth covered by the PAN sensor. However, we considered only 63 of them since the remaining three bands were zero throughout all the obtained data.

Although the sampling ratio of the satellite is 6, we generated data for $s=4$ because several state-of-the-art methods we compare with were not adaptable to different samplings. Since we have the spectral response provided by the PRISMA mission engineers, we were able to generate the PAN image from the HS data at the same resolution. According to the Wald's protocol \cite{wald1997fusion}, we consider the HS image as the reference one and generate the PAN image using the spectral response and the low-resolution HS image by downsampling by a factor of $4$. We cropped the reference images into patches of size $128\times 128$ and divided them into training, validation, and testing sets with ratios of 70$\%$, 15$\%$, and 15$\%$, respectively.

Table \ref{tab:prisma} shows the quantitative metrics obtained for each fusion method on both validation and testing sets. Our approach achieves the best results (highlighted in bold) for all metrics except SSIM, where SRPPNN performs better. Overall, the proposed method ranks first, with SRPPN and AWFLN ranking second and third, respectively. Additionally, Figure \ref{fig:rank-metrics}a compares the rank positions in terms of PSNR between the validation and testing sets. Most methods perform similarly on both sets, except for LGTEUN and GPPNN. Our method achieves the best performance, which shows its superior generalization capability over the PRISMA dataset.

\begin{figure*}[p!]
    \centering
    \begin{tabular}{c@{\hskip 0.2em}c@{\hskip 0.2em}c@{\hskip 0.2em}c@{\hskip 0.2em}c}
        \includegraphics[ trim=5cm 1cm 3cm 7cm, clip, width=0.17\linewidth]{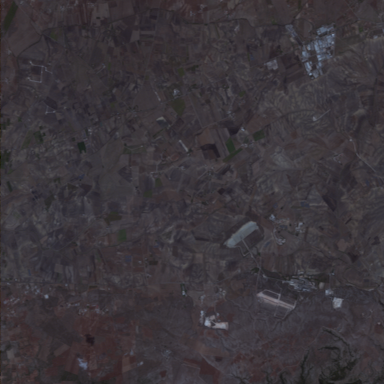} &
        \includegraphics[ trim=5cm 1cm 3cm 7cm, clip, width=0.17\linewidth]{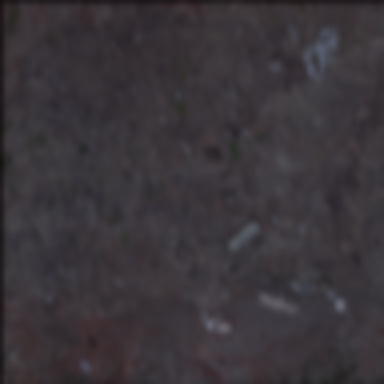} &
        \includegraphics[ trim=5cm 1cm 3cm 7cm, clip, width=0.17\linewidth]{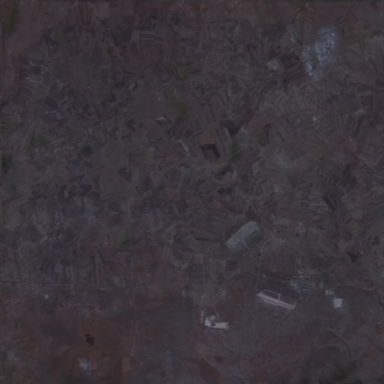} &
        \includegraphics[ trim=5cm 1cm 3cm 7cm, clip, width=0.17\linewidth]{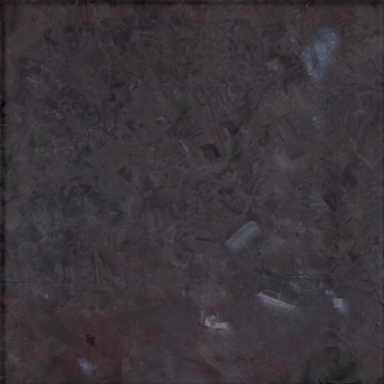} &
        \includegraphics[ trim=5cm 1cm 3cm 7cm, clip, width=0.17\linewidth]{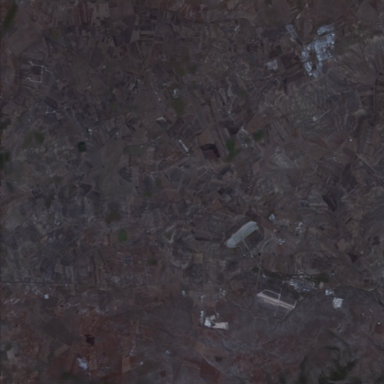}
        \\
        \includegraphics[ trim=5cm 1cm 3cm 7cm, clip, width=0.17\linewidth]{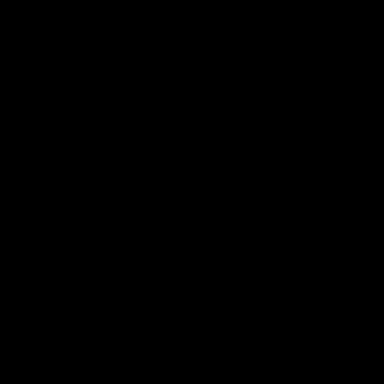} & 
        \includegraphics[ trim=5cm 1cm 3cm 7cm, clip, width=0.17\linewidth]{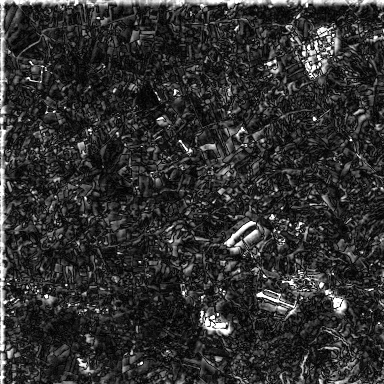}&
        \includegraphics[ trim=5cm 1cm 3cm 7cm, clip, width=0.17\linewidth]{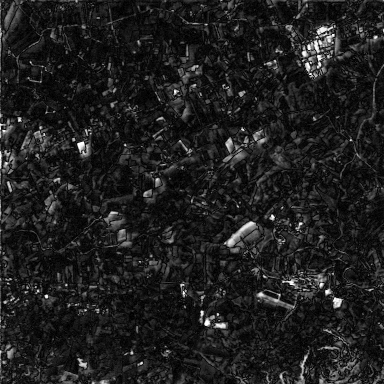}&
        \includegraphics[ trim=5cm 1cm 3cm 7cm, clip, width=0.17\linewidth]{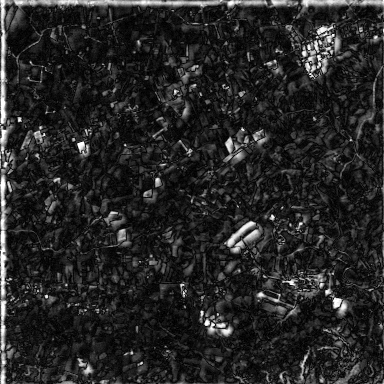}&
        \includegraphics[ trim=5cm 1cm 3cm 7cm, clip, width=0.17\linewidth]{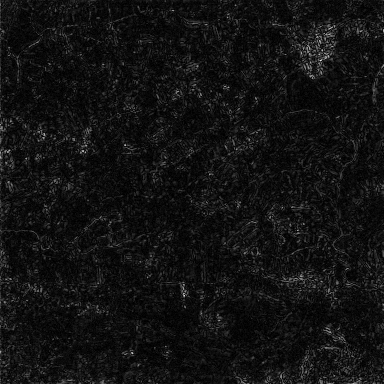}
        \\
        Reference & 
        Bicubic&
        Brovey &
        GSA &
        PanNet
        \\
        \includegraphics[ trim=5cm 1cm 3cm 7cm, clip, width=0.17\linewidth]{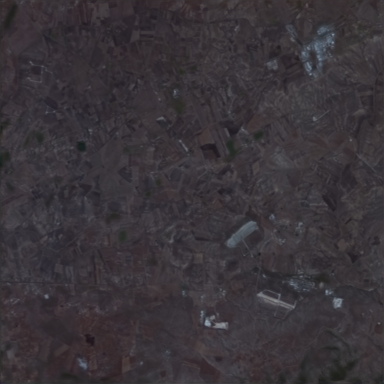}&
        \includegraphics[ trim=5cm 1cm 3cm 7cm, clip, width=0.17\linewidth]{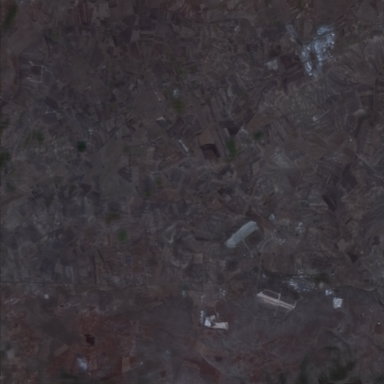} &
        \includegraphics[ trim=5cm 1cm 3cm 7cm, clip, width=0.17\linewidth]{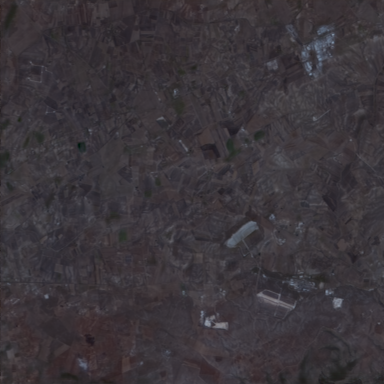} &
        \includegraphics[ trim=5cm 1cm 3cm 7cm, clip, width=0.17\linewidth]{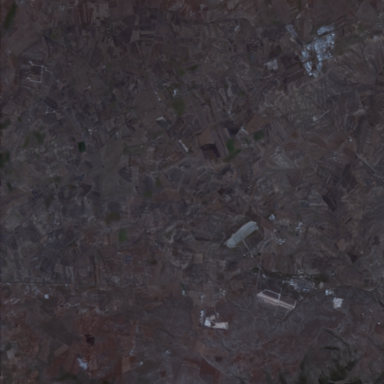}&
        \includegraphics[ trim=5cm 1cm 3cm 7cm, clip, width=0.17\linewidth]{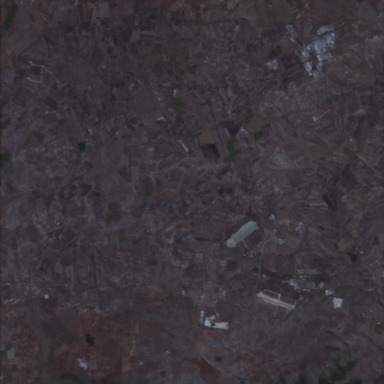}
        \\
        \includegraphics[ trim=5cm 1cm 3cm 7cm, clip, width=0.17\linewidth]{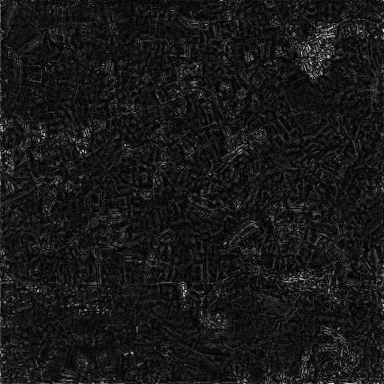}&
        \includegraphics[ trim=5cm 1cm 3cm 7cm, clip, width=0.17\linewidth]{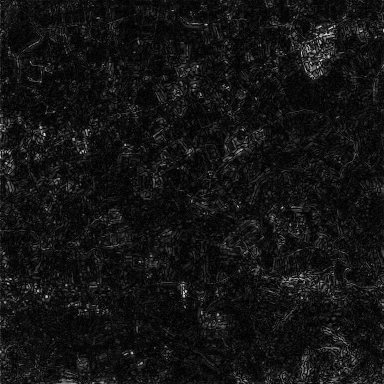} &
        \includegraphics[ trim=5cm 1cm 3cm 7cm, clip, width=0.17\linewidth]{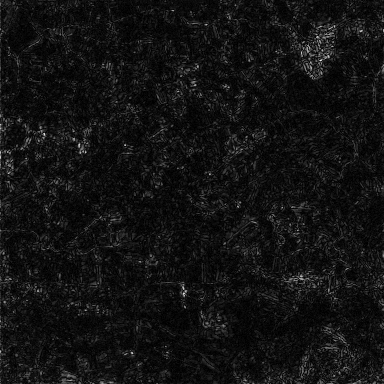} &
        \includegraphics[ trim=5cm 1cm 3cm 7cm, clip, width=0.17\linewidth]{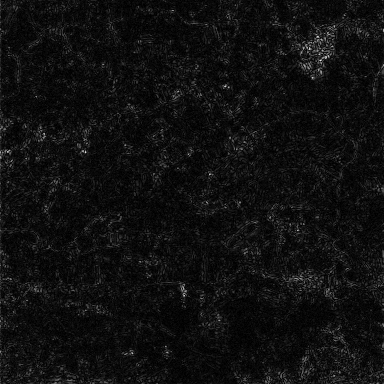}&
        \includegraphics[ trim=5cm 1cm 3cm 7cm, clip, width=0.17\linewidth]{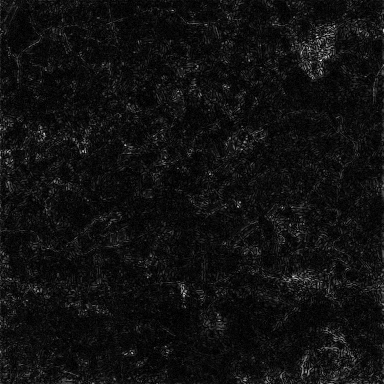}
        \\
        MSDCNN&
        DiCNN &
        FusionNet & 
        SRPPNN&
        HSIT
        \\
        \includegraphics[ trim=5cm 1cm 3cm 7cm, clip, width=0.17\linewidth]{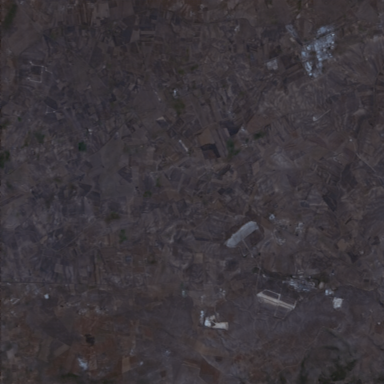} &
        \includegraphics[ trim=5cm 1cm 3cm 7cm, clip, width=0.17\linewidth]{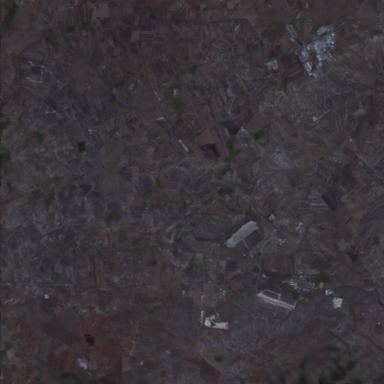}&
        \includegraphics[ trim=5cm 1cm 3cm 7cm, clip, width=0.17\linewidth]{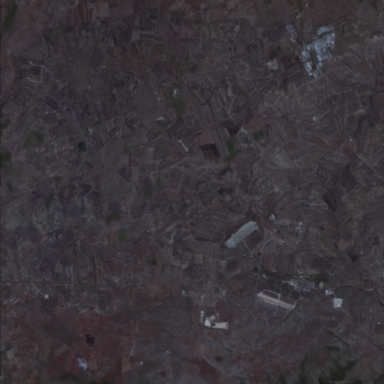} &
        \includegraphics[ trim=5cm 1cm 3cm 7cm, clip, width=0.17\linewidth]{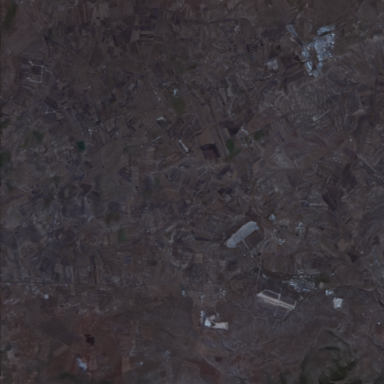}&
        \includegraphics[ trim=5cm 1cm 3cm 7cm, clip, width=0.17\linewidth]{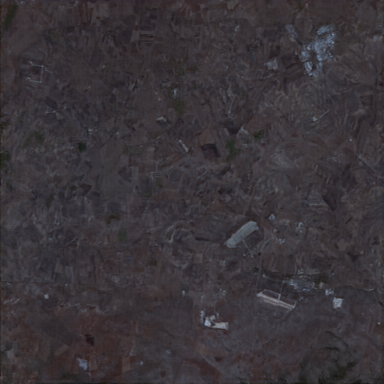} 
        \\
        \includegraphics[ trim=5cm 1cm 3cm 7cm, clip, width=0.17\linewidth]{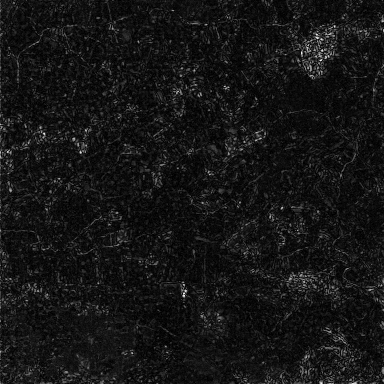} &
        \includegraphics[ trim=5cm 1cm 3cm 7cm, clip, width=0.17\linewidth]{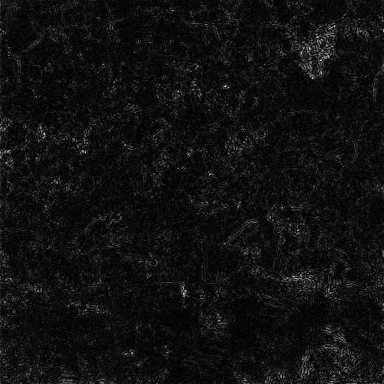}&
        \includegraphics[ trim=5cm 1cm 3cm 7cm, clip, width=0.17\linewidth]{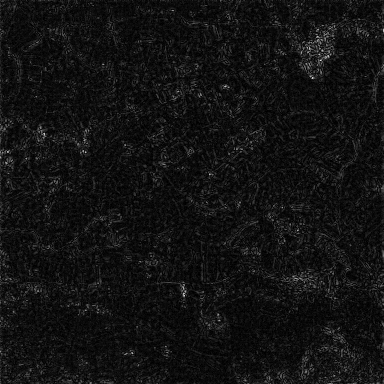} &
        \includegraphics[ trim=5cm 1cm 3cm 7cm, clip, width=0.17\linewidth]{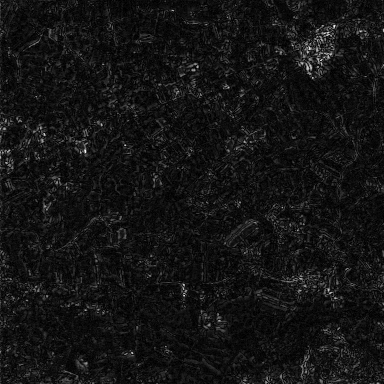}&
        \includegraphics[ trim=5cm 1cm 3cm 7cm, clip, width=0.17\linewidth]{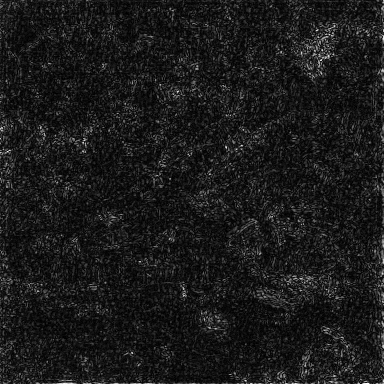} 
        \\
        PanFormer&
        AWFLN &
        MHFNet& 
        GPPNN&
        MMNet
        \\
        \includegraphics[ trim=5cm 1cm 3cm 7cm, clip, width=0.17\linewidth]{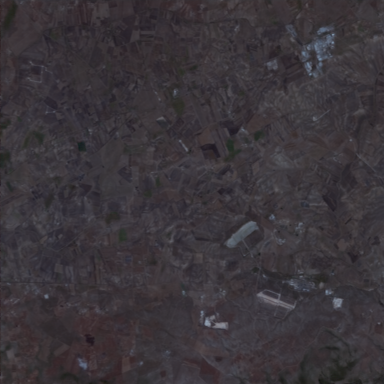}&
        \includegraphics[ trim=5cm 1cm 3cm 7cm, clip, width=0.17\linewidth]{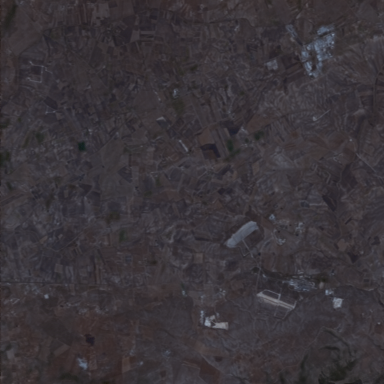}&
        \includegraphics[ trim=5cm 1cm 3cm 7cm, clip, width=0.17\linewidth]{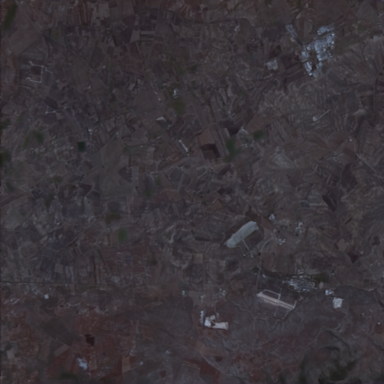}&
        \includegraphics[ trim=5cm 1cm 3cm 7cm, clip, width=0.17\linewidth]{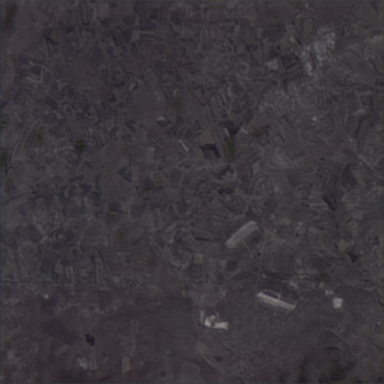} &
        \includegraphics[ trim=5cm 1cm 3cm 7cm, clip, width=0.17\linewidth]{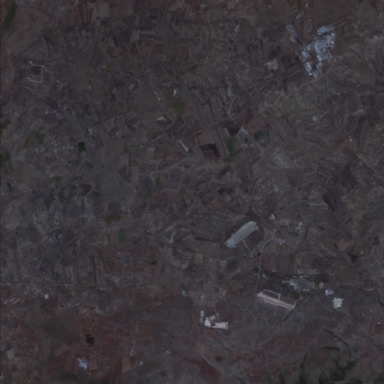}
        \\
        \includegraphics[ trim=5cm 1cm 3cm 7cm, clip, width=0.17\linewidth]{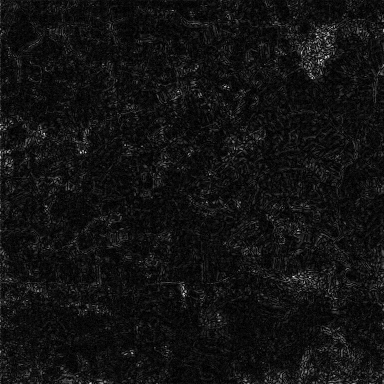}&
        \includegraphics[ trim=5cm 1cm 3cm 7cm, clip, width=0.17\linewidth]{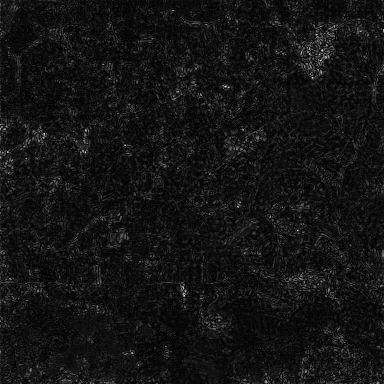}&
        \includegraphics[ trim=5cm 1cm 3cm 7cm, clip, width=0.17\linewidth]{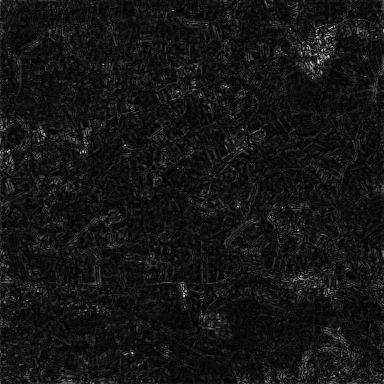}&
        \includegraphics[ trim=5cm 1cm 3cm 7cm, clip, width=0.17\linewidth]{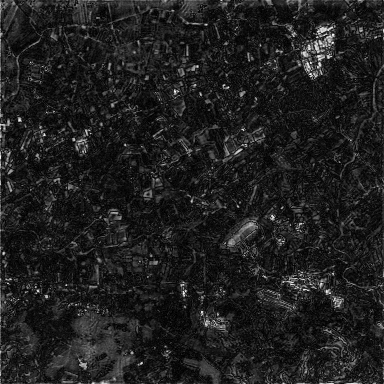} &
        \includegraphics[ trim=5cm 1cm 3cm 7cm, clip, width=0.17\linewidth]{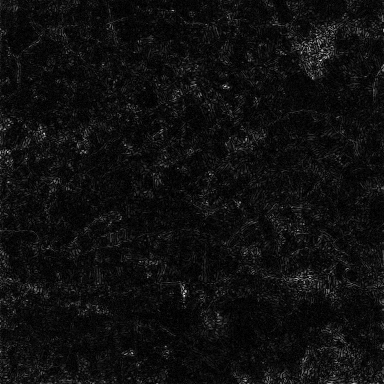}
        \\
        HyFPan&
        S2DBPN &
        LGTEUN & 
        UTeRM\_CNN&
        Ours
        \\
    \end{tabular}
    \caption{Visual comparison of the fusion methods on a cropped image from the PRISMA testing set. The 33rd, 45th, and 55th spectral bands are used in place of the RGB channels. The error maps are computed as the mean of the absolute value differences between the fused and reference images across all channels, and have been clipped and rescaled to highlight the errors. Furthermore, white balance and gamma correction are applied to each channel of the fused images for better visualization. Our method exhibits the highest visual quality, as evidenced by the error map containing less information.}
    \label{fig:prisma-results}
\end{figure*}

Figure \ref{fig:prisma-results} shows crops of the fused results produced by each method on a sample from the testing set. For visualization purposes, the 33rd, 45th, and 55th spectral bands are used in place of the RGB channels. Additionally, we provide the error maps, which are computed as the mean of the absolute value differences between the fused and reference images across all channels, and have been clipped and rescaled to highlight the errors. Consistent with the numerical results, our method exhibits the highest visual quality, as evidenced by the error map containing less information.

\subsection{Experiments on QuickBird}

QuickBird, which was in orbit from 2001 to 2015, was a pioneering commercial Earth observation satellite known for its very high-resolution imagery. It was equipped with four MS bands (blue, green, red, and near-infrared) along with a high-resolution PAN image. The MS data covers the nadir region (450-890 nm) with a spatial resolution of 2.4 m, while the PAN image has a resolution of 60 cm. We use the dataset available from a public repository\footnote{\url{https://github.com/liangjiandeng/PanCollection}}, where it has already been divided into training, validation and testing sets.

In Table \ref{tab:QuickBird}, we present the quantitative results for each fusion technique. Our method is the only one that ranks among the top three across all metrics for both validation and testing sets. Interestingly, although SRPPNN and AWFLN obtained the second and third best results on PRISMA, they are less competitive on Quickbird. In contrast, our proposal demonstrates consistent adaptability and robustness to different types of data. Figure \ref{fig:rank-metrics}b shows that S2DBPN and LTGTEUN, which achieve the best metrics in validation, do not generalize well to the testing set. MMNet, MHFNet and our method exhibit superior performance in this regard.

\begin{figure*}[p!]
    \centering
    \begin{tabular}{c@{\hskip 0.2em}c@{\hskip 0.2em}c@{\hskip 0.2em}c@{\hskip 0.2em}c}
        \includegraphics[ trim=1cm 4cm 3cm 0cm, clip, width=0.17\linewidth]{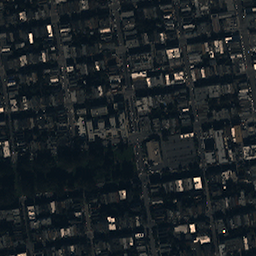} &
        \includegraphics[ trim=1cm 4cm 3cm 0cm, clip, width=0.17\linewidth]{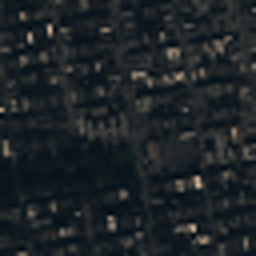} &
        \includegraphics[ trim=1cm 4cm 3cm 0cm, clip, width=0.17\linewidth]{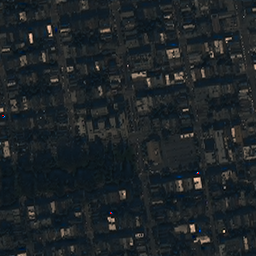} &
        \includegraphics[ trim=1cm 4cm 3cm 0cm, clip, width=0.17\linewidth]{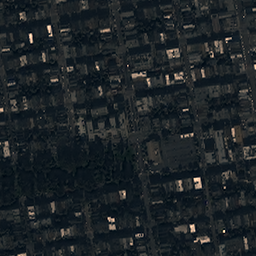} &
        \includegraphics[ trim=1cm 4cm 3cm 0cm, clip, width=0.17\linewidth]{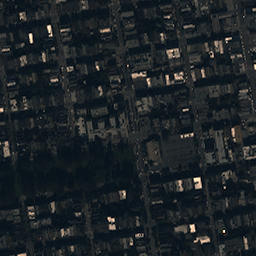}
        \\
        \includegraphics[ trim=1cm 4cm 3cm 0cm, clip, width=0.17\linewidth]{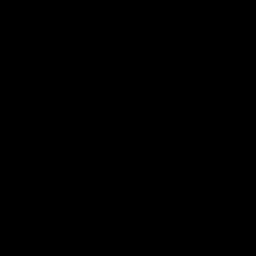} & 
        \includegraphics[ trim=1cm 4cm 3cm 0cm, clip, width=0.17\linewidth]{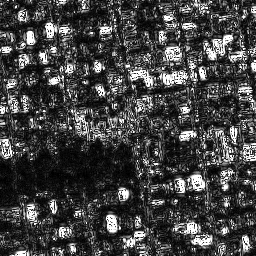}&
        \includegraphics[ trim=1cm 4cm 3cm 0cm, clip, width=0.17\linewidth]{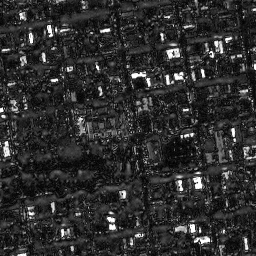}&
        \includegraphics[ trim=1cm 4cm 3cm 0cm, clip, width=0.17\linewidth]{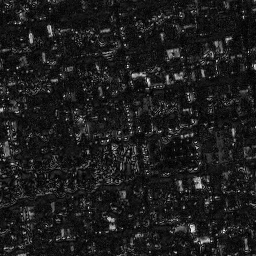}&
        \includegraphics[ trim=1cm 4cm 3cm 0cm, clip, width=0.17\linewidth]{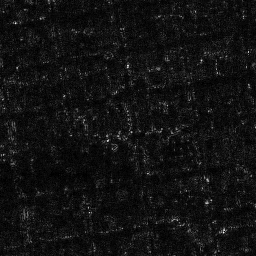}
        \\
        Reference & 
        Bicubic&
        Brovey &
        GSA &
        PanNet
        \\
        \includegraphics[ trim=1cm 4cm 3cm 0cm, clip, width=0.17\linewidth]{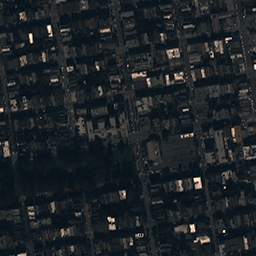}&
        \includegraphics[ trim=1cm 4cm 3cm 0cm, clip, width=0.17\linewidth]{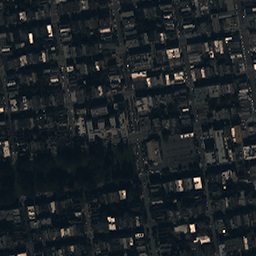} &
        \includegraphics[ trim=1cm 4cm 3cm 0cm, clip, width=0.17\linewidth]{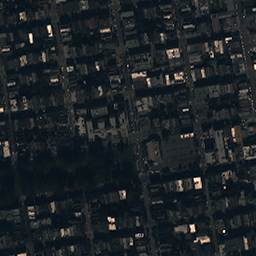} &
        \includegraphics[ trim=1cm 4cm 3cm 0cm, clip, width=0.17\linewidth]{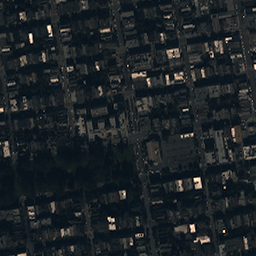}&
        \includegraphics[ trim=1cm 4cm 3cm 0cm, clip, width=0.17\linewidth]{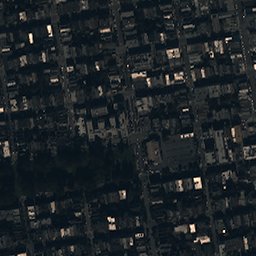}
        \\
        \includegraphics[ trim=1cm 4cm 3cm 0cm, clip, width=0.17\linewidth]{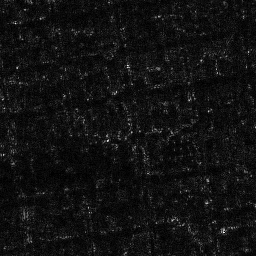}&
        \includegraphics[ trim=1cm 4cm 3cm 0cm, clip, width=0.17\linewidth]{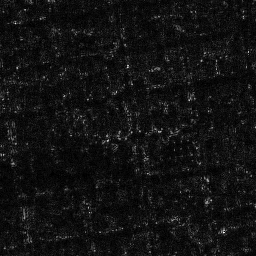} &
        \includegraphics[ trim=1cm 4cm 3cm 0cm, clip, width=0.17\linewidth]{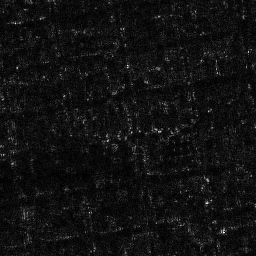} &
        \includegraphics[ trim=1cm 4cm 3cm 0cm, clip, width=0.17\linewidth]{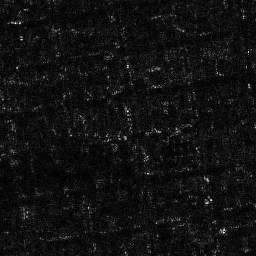}&
        \includegraphics[ trim=1cm 4cm 3cm 0cm, clip, width=0.17\linewidth]{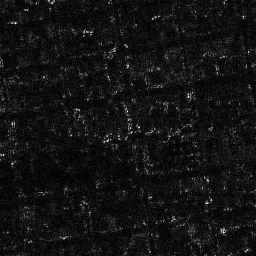}
        \\
        MSDCNN&
        DiCNN &
        FusionNet & 
        SRPPNN&
        HSIT
        \\
        \includegraphics[ trim=1cm 4cm 3cm 0cm, clip, width=0.17\linewidth]{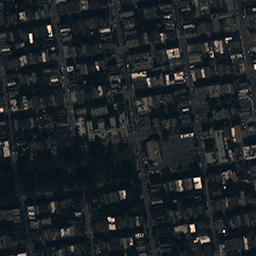} &
        \includegraphics[ trim=1cm 4cm 3cm 0cm, clip, width=0.17\linewidth]{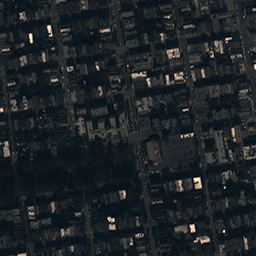}&
        \includegraphics[ trim=1cm 4cm 3cm 0cm, clip, width=0.17\linewidth]{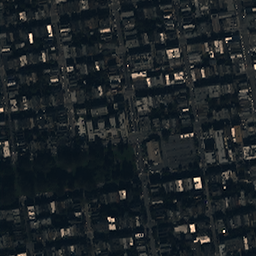} &
        \includegraphics[ trim=1cm 4cm 3cm 0cm, clip, width=0.17\linewidth]{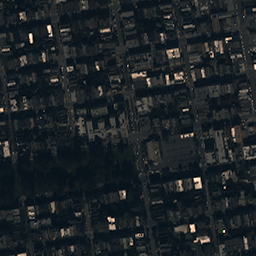}&
        \includegraphics[ trim=1cm 4cm 3cm 0cm, clip, width=0.17\linewidth]{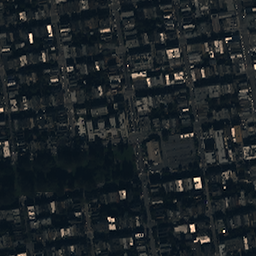} 
        \\
        \includegraphics[ trim=1cm 4cm 3cm 0cm, clip, width=0.17\linewidth]{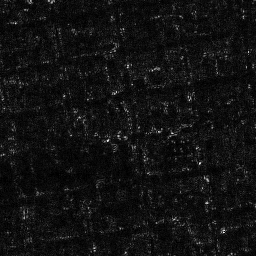} &
        \includegraphics[ trim=1cm 4cm 3cm 0cm, clip, width=0.17\linewidth]{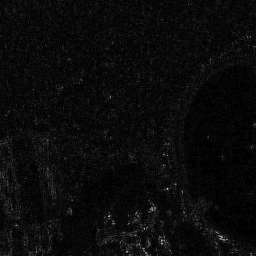}&
        \includegraphics[ trim=1cm 4cm 3cm 0cm, clip, width=0.17\linewidth]{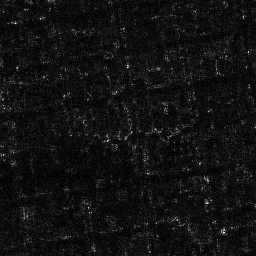} &
        \includegraphics[ trim=1cm 4cm 3cm 0cm, clip, width=0.17\linewidth]{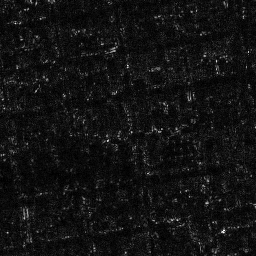}&
        \includegraphics[ trim=1cm 4cm 3cm 0cm, clip, width=0.17\linewidth]{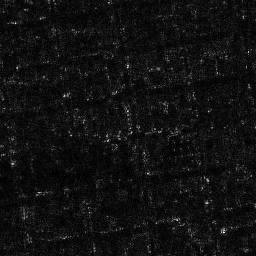} 
        \\
        PanFormer&
        AWFLN &
        MHFNet& 
        GPPNN&
        MMNet
        \\
        \includegraphics[ trim=1cm 4cm 3cm 0cm, clip, width=0.17\linewidth]{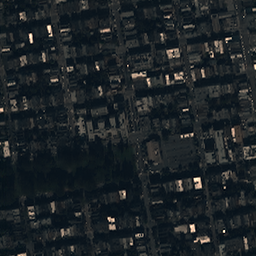}&
        \includegraphics[ trim=1cm 4cm 3cm 0cm, clip, width=0.17\linewidth]{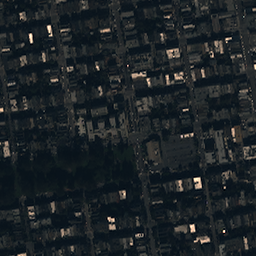}&
        \includegraphics[ trim=1cm 4cm 3cm 0cm, clip, width=0.17\linewidth]{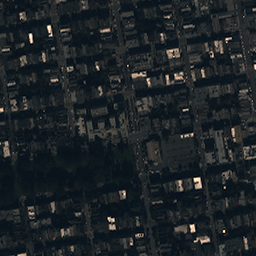}&
        \includegraphics[ trim=1cm 4cm 3cm 0cm, clip, width=0.17\linewidth]{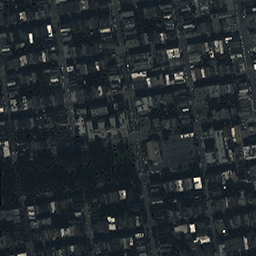} &
        \includegraphics[ trim=1cm 4cm 3cm 0cm, clip, width=0.17\linewidth]{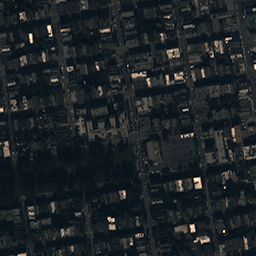}
        \\
        \includegraphics[ trim=1cm 4cm 3cm 0cm, clip, width=0.17\linewidth]{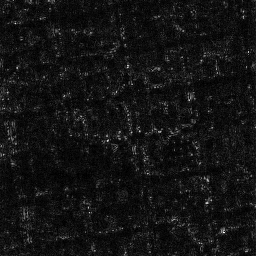}&
        \includegraphics[ trim=1cm 4cm 3cm 0cm, clip, width=0.17\linewidth]{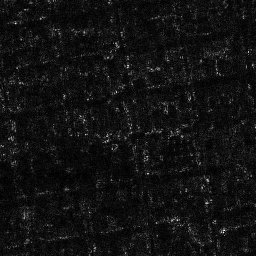}&
        \includegraphics[ trim=1cm 4cm 3cm 0cm, clip, width=0.17\linewidth]{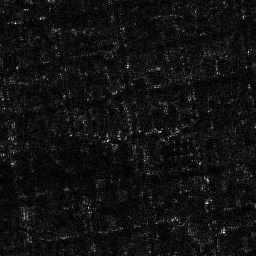}&
        \includegraphics[ trim=1cm 4cm 3cm 0cm, clip, width=0.17\linewidth]{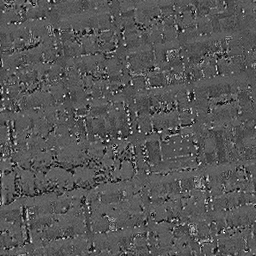} &
        \includegraphics[ trim=1cm 4cm 3cm 0cm, clip, width=0.17\linewidth]{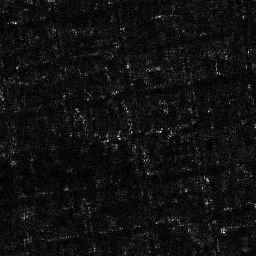}
        \\
        HyFPan&
        S2DBPN &
        LGTEUN & 
        UTeRM\_CNN&
        Ours
        \\
    \end{tabular}

\caption{Visual comparison of the fusion methods on a cropped image from the Quickbird testing set. The 3rd, 2nd, and 1st spectral bands are used as RGB channels. The error maps are computed as the mean of the absolute value differences between the fused and reference images across all channels, and have been clipped and rescaled to highlight the errors. Furthermore, white balance and gamma correction are applied to each channel of the fused images for better visualization. Although discerning significant differences between various fused images seems challenging, the error maps associated with MMNet and our method are less pronounced.}

    \label{fig:quickbird-results}
\end{figure*}

Figure \ref{fig:quickbird-results} presents crops of the fused images and the corresponding error maps generated by each method on a testing sample. The 3rd, 2nd and 1st spectral bands are used as RGB channels. Visually, it is difficult to discern significant differences between various fused images. However, the error maps associated with MMNet and our method are less pronounced, which indicates better spatial and spectral reconstruction.

\subsection{Experiments on WorldView2}

The WorldView2 sensor was designed to operate within the visible and near-infrared spectrum. This satellite provides a MS image that comprises eight spectral bands (coastal, blue, green, yellow, red, red edge, near-infrared region 1, and near-infrared region 2) along with a PAN image. The spatial resolutions are 1.84 m for MS data and 0.46 m for PAN. 

We generated the WorldView2 dataset from data obtained through the ESA portal\footnote{\url{https://earth.esa.int/eogateway/catalog/worldview-2-full-archive-and-tasking}\label{ref-world}}. We cropped the original MS images into non-overlapping tiles of $128\times 128$ pixels and the original PAN images into non-overlapping tiles of $512\times 512$ pixels. Then, we applied Wald's Protocol \cite{wald1997fusion} to these crops to generate labeled data for evaluation. We divided the resulting crops into training, validation, and testing sets with ratios of 70$\%$, 15$\%$, and 15$\%$, respectively. 

Table \ref{tab:WorldView2} presents the quantitative results for the validation and testing sets. The proposed model outperforms all others across all metrics. The second and third best methods are MMNet and SRPPNN, respectively. It is noteworthy that MMNet yield some of the poorest results on the PRISMA dataset among pure deep learning and unfolded techniques, while SRPPNN was less competitive on Quickbird. Figure \ref{fig:rank-metrics}c compares the rank positions in terms of PSNR between the validation and testing sets. Except for PanFormer and LGTEUN, all methods perform similarly on both sets, with our proposal being the highest ranked.

\begin{figure*}[p!]
    \centering
    \begin{tabular}{c@{\hskip 0.2em}c@{\hskip 0.2em}c@{\hskip 0.2em}c@{\hskip 0.2em}c}
        \includegraphics[ trim=1.5cm 0cm 0cm 1.5cm, clip, width=0.17\linewidth]{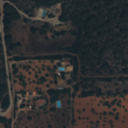} &
        \includegraphics[ trim=1.5cm 0cm 0cm 1.5cm, clip, width=0.17\linewidth]{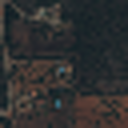} &
        \includegraphics[ trim=1.5cm 0cm 0cm 1.5cm, clip, width=0.17\linewidth]{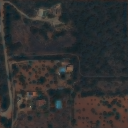} &
        \includegraphics[ trim=1.5cm 0cm 0cm 1.5cm, clip, width=0.17\linewidth]{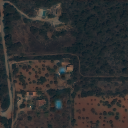} &
        \includegraphics[ trim=1.5cm 0cm 0cm 1.5cm, clip, width=0.17\linewidth]{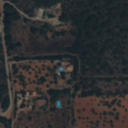}
        \\
        \includegraphics[ trim=1.5cm 0cm 0cm 1.5cm, clip, width=0.17\linewidth]{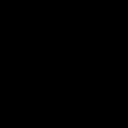} & 
        \includegraphics[ trim=1.5cm 0cm 0cm 1.5cm, clip, width=0.17\linewidth]{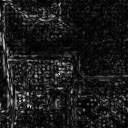}&
        \includegraphics[ trim=1.5cm 0cm 0cm 1.5cm, clip, width=0.17\linewidth]{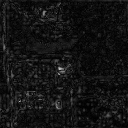}&
        \includegraphics[ trim=1.5cm 0cm 0cm 1.5cm, clip, width=0.17\linewidth]{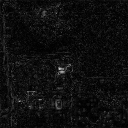}&
        \includegraphics[ trim=1.5cm 0cm 0cm 1.5cm, clip, width=0.17\linewidth]{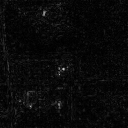}
        \\
        Reference & 
        Bicubic&
        Brovey &
        GSA &
        PanNet
        \\
        \includegraphics[ trim=1.5cm 0cm 0cm 1.5cm, clip, width=0.17\linewidth]{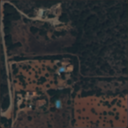}&
        \includegraphics[ trim=1.5cm 0cm 0cm 1.5cm, clip, width=0.17\linewidth]{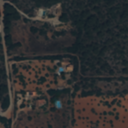} &
        \includegraphics[ trim=1.5cm 0cm 0cm 1.5cm, clip, width=0.17\linewidth]{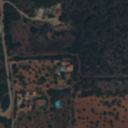} &
        \includegraphics[ trim=1.5cm 0cm 0cm 1.5cm, clip, width=0.17\linewidth]{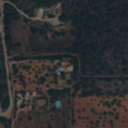}&
        \includegraphics[ trim=1.5cm 0cm 0cm 1.5cm, clip, width=0.17\linewidth]{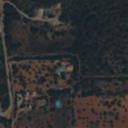}
        \\
        \includegraphics[ trim=1.5cm 0cm 0cm 1.5cm, clip, width=0.17\linewidth]{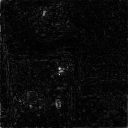}&
        \includegraphics[ trim=1.5cm 0cm 0cm 1.5cm, clip, width=0.17\linewidth]{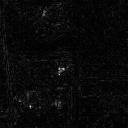} &
        \includegraphics[ trim=1.5cm 0cm 0cm 1.5cm, clip, width=0.17\linewidth]{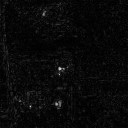} &
        \includegraphics[ trim=1.5cm 0cm 0cm 1.5cm, clip, width=0.17\linewidth]{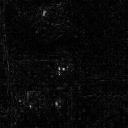}&
        \includegraphics[ trim=1.5cm 0cm 0cm 1.5cm, clip, width=0.17\linewidth]{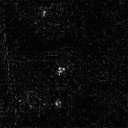}
        \\
        MSDCNN&
        DiCNN &
        FusionNet & 
        SRPPNN&
        HSIT
        \\
        \includegraphics[ trim=1.5cm 0cm 0cm 1.5cm, clip, width=0.17\linewidth]{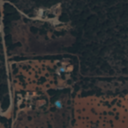} &
        \includegraphics[ trim=1.5cm 0cm 0cm 1.5cm, clip, width=0.17\linewidth]{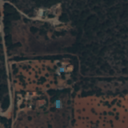}&
        \includegraphics[ trim=1.5cm 0cm 0cm 1.5cm, clip, width=0.17\linewidth]{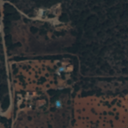} &
        \includegraphics[ trim=1.5cm 0cm 0cm 1.5cm, clip, width=0.17\linewidth]{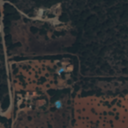}&
        \includegraphics[ trim=1.5cm 0cm 0cm 1.5cm, clip, width=0.17\linewidth]{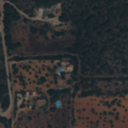} 
        \\
        \includegraphics[ trim=1.5cm 0cm 0cm 1.5cm, clip, width=0.17\linewidth]{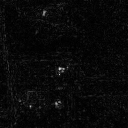} &
        \includegraphics[ trim=1.5cm 0cm 0cm 1.5cm, clip, width=0.17\linewidth]{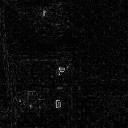}&
        \includegraphics[ trim=1.5cm 0cm 0cm 1.5cm, clip, width=0.17\linewidth]{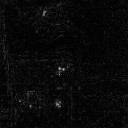} &
        \includegraphics[ trim=1.5cm 0cm 0cm 1.5cm, clip, width=0.17\linewidth]{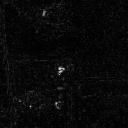}&
        \includegraphics[ trim=1.5cm 0cm 0cm 1.5cm, clip, width=0.17\linewidth]{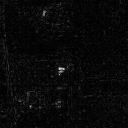} 
        \\
        PanFormer&
        AWFLN &
        MHFNet& 
        GPPNN&
        MMNet
        \\
        \includegraphics[ trim=1.5cm 0cm 0cm 1.5cm, clip, width=0.17\linewidth]{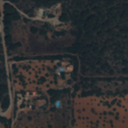}&
        \includegraphics[ trim=1.5cm 0cm 0cm 1.5cm, clip, width=0.17\linewidth]{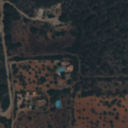}&
        \includegraphics[ trim=1.5cm 0cm 0cm 1.5cm, clip, width=0.17\linewidth]{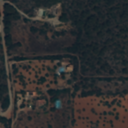}&
        \includegraphics[ trim=1.5cm 0cm 0cm 1.5cm, clip, width=0.17\linewidth]{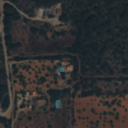} &
        \includegraphics[ trim=1.5cm 0cm 0cm 1.5cm, clip, width=0.17\linewidth]{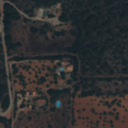}
        \\
        \includegraphics[ trim=1.5cm 0cm 0cm 1.5cm, clip, width=0.17\linewidth]{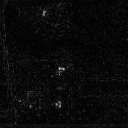}&
        \includegraphics[ trim=1.5cm 0cm 0cm 1.5cm, clip, width=0.17\linewidth]{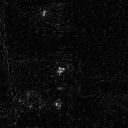}&
        \includegraphics[ trim=1.5cm 0cm 0cm 1.5cm, clip, width=0.17\linewidth]{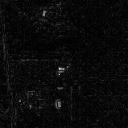}&
        \includegraphics[ trim=1.5cm 0cm 0cm 1.5cm, clip, width=0.17\linewidth]{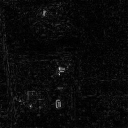} &
        \includegraphics[ trim=1.5cm 0cm 0cm 1.5cm, clip, width=0.17\linewidth]{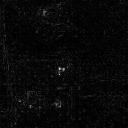}
        \\
        HyFPan&
        S2DBPN &
        LGTEUN & 
        UTeRM\_CNN&
        Ours
        \\
    \end{tabular}
    \caption{Visual comparison of the fusion methods on a cropped image from the WorldView2 testing set. The 4th, 2nd, and 1st spectral bands are used in place of the RGB channels. The error maps are computed as the mean of the absolute value differences between the fused and reference images across all channels, and have been clipped and rescaled to highlight the errors. Furthermore, white balance and gamma correction are applied to each channel of the fused images for better visualization. Our method provides superior visual results in terms of both spectral consistency and geometric accuracy. For instance, unlike our approach, all other methods exhibit a noticeable drooling effect on the swimming pools.}

    \label{fig:WorldView2}
\end{figure*}

Figure \ref{fig:WorldView2} shows the outcomes of each method and the corresponding error maps. The 4th, 2nd, and 1st spectral bands are used in place of the RGB channels. We observe that our method provides superior visual results in terms of spectral consistency and geometric accuracy. For instance, unlike our approach, all other fusion techniques exhibit a noticeable drooling effect on the swimming pools, causing the water to appear as though it extends beyond the boundaries of the structure.

\begin{sidewaystable*}
\centering
\caption{Quantitative metrics obtained for each fusion method on the PRISMA validation and testing sets. The best values are highlighted in bold, the second best in blue and the third best in red. Our approach achieves the best results for all metrics except SSIM, where SRPPNN performs better.}

\begin{tabular}{l |c c |c c |c c |c c |c c}
 & \multicolumn{2}{c}{ERGAS$\downarrow$} & \multicolumn{2}{c}{PSNR$\uparrow$} & \multicolumn{2}{c}{SSIM$\uparrow$} & \multicolumn{2}{c}{Q$2^n\uparrow$}& \multicolumn{2}{c}{SAM$\downarrow$} \\
 & validation & test& validation & test & validation & test  & validation &test  & validation & test \\
\hline
Bicubic &  105,89& 81,74&  29,79& 31,66& 0,7273& 0,7951& 0,205& 0,204& 4,35& 3,61\\
Brovey & 59,13& 47,23& 34,98& 36,45& 0,8954& 0,9205& 0,600& 0,580& 4,70&3,83\\
Wavelet &  97,19& 75,40&  30,54& 32,38& 0,8036& 0,8513& 0,554& 0,563& 5,29& 4,38\\
 PCA& 285.52 & 238.54& 20.55 & 22.94 & 0.3312& 0.5554& 0.159& 0.182& 30,79&27,96\\
 IHS& 68.19 & 55.09& 30.84 & 35.16 & 0.8640& 0.8972& 0.662& 0.681& 5,94&4,53\\
GSA & 89,53& 61,23& 31,16& 34,14& 0,8593& 0,9091& 0,531& 0,577&4,81&  4,15\\
CNMF &  86,24& 63,04&  31,45& 33,92&  0,8217&  0,8792& 0,381& 0,436& 4,28& 3,69\\
 \hline
PanNet& 29,07& 24,35& 40,61& 42,06& 0,9494& 0,9622& 0,791&0,788& 3,04& 2,46\\
MSDCNN& 35,21& 32,01& 38,95& 39,65& 0,9435& 0,9575& 0,761&0,759& 3,91& 3,77\\
DiCNN& 30,72& 26,50& 40,09& 41,34& 0,9395& 0,9531& 0,770&0,755& 3,22& 2,68\\
FusionNet& 27,91& 23,96& 40,97& 42,20& 0,9491& 0,9608& 0,796&0,791& 2,96& 2,47\\
SRPPNN& {\color{blue} 25,62}& {\color{blue} 21,44}& {\color{blue} 41,76}& \textbf{43,20}& \textbf{0,9570}& \textbf{0,9676}& {\color{blue} 0,816}&\textbf{0,816}& {\color{blue} 2,74}& {\color{blue} 2,25}\\
HSIT& 29,00 & 24,25 & 40,69 & 42,06 & 0,9503& 0,9624&  0,790& 0,785 & 3,31& 2,70\\
PanFormer& 35,13& 31,11& 39,11& 39,94& 0,9385& 0,9519&  0,725& 0,739& 3,98& 3,535\\
AWFLN& {\color{red} 26,19}& {\color{red} 22,33}& {\color{red} 41,54}& {\color{blue} 42,85}& {\color{red} 0,9549}& {\color{red} 0,9647}& {\color{blue} 0,816}&{\color{blue} 0,803}& {\color{red} 2,78}& {\color{red}2,32}\\
\hline
MHFNet& 27,47& 22,71& 41,19& {\color{red} 42,67}& 0,9537& 0,9654& 0,804&{\color{red} 0,798}& 3,04& 2,46\\
GPPNN& 28,15& 25,72& 40,92& 41,58& 0,9515& 0,9597& 0,793&0,776& 3,13& 2,71\\
MMNet& 67,53& 44,11& 36,47&37,95&0,8857& 0,9284& 0,717&0,712& 7,99& 5,43\\
HyFPan& 28,30& 24,48& 40,86& 42,07& 0,9500& 0,9630& {\color{red} 0,806}&0,791& 3,13& 2,759\\
S2DBPN& 29,68& 25,52& 40,51& 41,63& 0,9495& 0,9630& 0,792&0,786& 3,43& 2,98\\
LGTEUN& 27,22& 28,62& 41,25& 40,79& 0,9534& 0,9581&  0,737& 0,743& 2,93& 3,405\\
UTeRM\_CNN& 47,82& 40,20& 36,23& 37,61& 0,8910& 0,9204& 0,614&0,616& 4,75& 3,695\\
\hline
Ours& \textbf{25,36}& \textbf{21,41}& \textbf{41,80}& \textbf{43,20}& {\color{blue} 0,9565}& {\color{blue} 0,9673}& \textbf{0,831}& \textbf{0,816}& \textbf{2,67}& \textbf{2,23}\\
\hline
\end{tabular}
\label{tab:prisma}
\end{sidewaystable*}

\begin{sidewaystable*}
\centering
\caption{Quantitative metrics obtained for each fusion technique on the Quickbird validation and testing sets. The best values are highlighted in bold, the second best in blue and the third best in red. Our method is the only one that places all metrics, both in validation and testing, among the top three.}

\begin{tabular}{l  |c c |c c |c c |c c |c c}
 & \multicolumn{2}{c}{ERGAS$\downarrow
$} & \multicolumn{2}{c}{PSNR$\uparrow
$} & \multicolumn{2}{c}{SSIM$\uparrow
$} & \multicolumn{2}{c}{Q$2^n$$\uparrow
$}& \multicolumn{2}{c}{SAM$\downarrow
$} \\
 & validation & test  & validation & test & validation & test  & validation &test  & validation & test \\
 \hline
 
Bicubic & 169,69 & 165,98& 28,44 & 28,44& 0,7275& 0,7271& 0,396& 0,498& 20,40 &7,82\\
Brovey & 101,33 & 97,92& 32,99& 32,98& 0,9088& 0,9083& 0,696& 0,789& 11,39&8,12\\
Wavelet & 87,87 &86,20& 34,04 &34,04& 0,9074&0,9076&0,867&0,872& 10,42 &8,85\\
PCA& 156,43 & 126,71& 29,32 & 30,63& 0,8442& 0,8783& 0,551 & 0,642& 19,81 &15,47\\
IHS& 81,25 & 99,76& 32,81 & 32,82& 0,9146& 0,9035& 0,707 & 0,763& 8,55&8,41\\
GSA & 73,81 &72,64& 35,53 &35,52& 0,9325&0,9321&0,855&0,870& 8,61 &7,54\\
CNMF & 71,62 &69,54& 35,85 &35,88& 0,9413&0,9438&0,783&0,854& 8,37 &6,57\\ 
 \hline
PanNet& 34,95& 41,26& 43,10& 40,41& 0,9782& 0,9749& 0,903&0,931& 4,14& 4,56\\
MSDCNN& 34,34& 40,79& 43,27& 40,52& 0,9789& 0,9756& 0,903&{\color{red}0,932}& 3,89& 4,48\\
DiCNN& 35,93& 42,80& 42,93& 40,10& 0,9770& 0,9734&0,895&0,928& 4,12& 4,66\\
FusionNet& 34,31& 41,13& 43,27& 40,44& 0,9789& 0,9753& 0,901&0,931& 4,02& 4,50\\
SRPPNN& 33,34& 41,15& 43,51& 40,45& 0,9800& 0,9755& {\color{blue} 0,907}&{\color{blue}0,933}& 3,99& 4,42\\
HSIT& 36,16 & 46,23 & 42,89 & 39,45 & 0,9778& 0,9711& 0,902 & 0,925 & 4,10& 4,68\\
PanFormer& 34,72& 43,43& 43,20& 35,20& 0,9785& 0,9733& 0,897& 0,929& 3,97&4,45\\
AWFLN& 34,07&  40,19&  43,34& 40,65& 0,9794& 0,9764& 0,904&{\color{blue}0,933}& 3,94& {\color{red} 4,37}\\
\hline
MHFNet& 34,73& {\color{blue}40,04}&  43,16& {\color{blue} 40,70}& 0,9790& {\color{red} 0,9765} & 0,903&\textbf{0,934}& 4,15& 4,41\\
GPPNN& 34,38& 45,52&  43,26& 39,57& 0,9791& 0,9711& 0,903&0,924& 3,90& 4,80\\
MMNet& 33,38& \textbf{39,72}& 43,50& \textbf{40,74}& 0,9801& \textbf{0,9769}& {\color{red} 0,906}&\textbf{0,934}& 3,86& {\color{blue} 4,32}\\
HyFPan& 39,90& 45,44&  41,99& 39,58& 0,9740& 0,9707& 0,891&0,923& 4,79& 4,98\\
S2DBPN& \textbf{31,91}& 44,26& \textbf{43,87}& 39,84& {\color{blue} 0,9814}& 0,9724& \textbf{0,910}&0,929& {\color{blue} 3,74}& 4,67\\
LGTEUN& {\color{blue} 31,99}& 40,69& {\color{blue} 43,85}& 35,38& \textbf{0,9815}& 0,9758& 0,904& {\color{red}0,932}& \textbf{3,68}&4,42\\
UTeRM\_CNN& 33,16& 104,82& 43,54& 29,98& 0,9803& 0,9043& 0,905&0,584& 3,80& 9,20\\
\hline
Ours&  {\color{red} 32,49}&  {\color{red}40,09}&  {\color{red} 43,72}&  {\color{red} 40,68}& {\color{red}0,9812}& {\color{blue} 0,9767} &  {\color{blue} 0,907}&  \textbf{0,934}&  {\color{red}3,76}&  \textbf{4,31}\\
\hline
\end{tabular}
\label{tab:QuickBird}
\end{sidewaystable*}

\begin{sidewaystable*}
\centering
\caption{Quantitative metrics obtained for each fusion method on the WorldView2 validation and testing sets. The best values are highlighted in bold, the second best in blue and the third best in red. Our approach achieves the best results for all metrics.}

\begin{tabular}{l |c c |c c |c c |c c |c c}
 & \multicolumn{2}{c}{ERGAS$\downarrow$}& \multicolumn{2}{c}{PSNR$\uparrow$} & \multicolumn{2}{c}{SSIM$\uparrow$} & \multicolumn{2}{c}{Q$2^n$$\uparrow$}& \multicolumn{2}{c}{SAM$\downarrow$} \\
 & validation & test & validation & test & validation & test  & validation &test  & validation & test \\
    \hline
    Bicubic& 51,89 & 52,79 & 31,84 & 31,57 & 0,7901& 0,7836& 0,542& 0,542& 3,16 & 3,22 \\
    Brovey& 38,02 & 38,70 & 34,51 & 34,26 & 0,8962& 0,8947& 0,786& 0,789& 3,39 & 3,45\\
    Wavelet& 43,89 & 44,64 & 33,26 & 33,01 & 0,8563& 0,8521& 0,786& 0,788& 3,42& 3,78 \\
    PCA& 134,29 & 130,72 & 23,26 & 23,58 & 0,7848& 0,8164& 0,438 & 0,463 & 18,12 & 17,89 \\
    IHS& 38,30 & 38,85 & 32,44 & 34,22 & 0,8895& 0,8864& 0,770 & 0,773 & 3,61& 3,66 \\
    GSA &  32,92& 33,31 &  35,79& 35,61 & 0,9209& 0,9187&  0,841&  0,845& 2,86& 2,72 \\
    CNMF& 35,48& 35,79& 35,14& 34,95& 0,9137& 0,9113& 0,776& 0,787& 2,89&2,95\\
    \hline
    PanNet& 20,03& 19,50&  40,12& 40,42& 0,9683& 0,9700& 0,931&0,928& 1,86& 1,82\\
    MSDCNN& 22,98& 22,44&  38,85& 39,13& 0,9647& 0,9666& 0,922&0,918& 2,10& 2,06\\
    DiCNN& 20,53& 20,06&  39,88& 40,16& 0,9668& 0,9684& 0,927&0,923& 1,92& 1,88\\
    FusionNet& 20,05& 19,35& 40,17& 40,51& 0,9692& 0,9707& 0,932& 0,928& 1,83& 1,79\\
    SRPPNN& {\color{red} 19,01}& {\color{red} 18,52}&  {\color{red} 40,60}& {\color{red} 40,89}& {\color{red} 0,9711}& {\color{red} 0,9726}& {\color{blue} 0,936}&{\color{blue}0,932}& {\color{blue} 1,76}& {\color{blue} 1,72}\\
    HSIT& 23,89 & 24,26 & 38,73 & 38,49 & 0,9607& 0,9590& 0,910 & 0,914 & 2,14& 2,16\\
    PanFormer& 19,42& 20,01& 40,49& 40,17& 0,9708& 0,9691& 0,928& {\color{blue} 0,932}& 1,80&1,84\\
    AWFLN& 21,75& 21,13& 39,41& 39,72& 0,9630& 0,9650& 0,920&0,918& 2,02& 1,97\\
    \hline
    MHFNet& 20,02& 19,52&  40,13& 40,42& 0,9689& 0,9704& 0,931&0,928& 1,89& 1,85\\
    GPPNN& 19,81& 19,24& 40,27& 40,58& 0,9698& 0,9716& {\color{red} 0,933}&0,930& 1,81& 1,76\\
    MMNet& {\color{blue} 18,86}& {\color{blue} 18,43}& {\color{blue} 40,67}& {\color{blue} 40,94}& {\color{blue} 0,9720}& {\color{blue} 0,9733}& {\color{blue} 0,936}&{\color{red} 0,931}& {\color{red} 1,77}& {\color{red}1,73}\\
    HyFPan& 19,55& 19,10&  40,30& 40,58& 0,9702& 0,9715& 0,931&0,927& 1,85& 1,82\\
    S2DBPN& 21,36& 20,81& 39,56& 39,85& 0,9641& 0,9658& 0,920&0,916& 2,04& 1,99\\
    LGTEUN& 21,69& 22,33& 39,50& 39,19& 0,9638& 0,9617& 0,913& 0,916& 2,04&2,09\\
    UTeRM\_CNN& 22,19& 21,59& 39,22& 39,52& 0,9623& 0,9643& 0,918&0,915& 2,05& 2,00\\
    \hline
    Ours& \textbf{18,74}& \textbf{18,26}& \textbf{40,74}& \textbf{41,03}& \textbf{0,9724} & \textbf{0,9737}& \textbf{0,939}& \textbf{0,935}& \textbf{1,72}& \textbf{1,68}\\
    \hline
\end{tabular}
\label{tab:WorldView2}
\end{sidewaystable*}

\subsection{Overall performance}

The proposed fusion method achieves the highest metrics on the Prisma and WorldView2 datasets. Furthermore, it is the only technique that ranks all metrics within the top three on Quickbird, where other methods exhibit inconsistent performance. Figure \ref{fig:rank-metrics} compares the average rank positions across all datasets between the validation and testing sets. These results highlight the superior performance of our approach, followed by SRPPNN.

\section{Ablation Study}\label{sec:ablation}

We have conducted several tests to identify the best configuration for our approach, evaluating its robustness and adaptability. We selected the most effective architecture for the MARNet from among various candidates. We also assessed various post-processing configurations. Finally, we analyzed the robustness of the proposed method to different sampling factors and noise.

\subsection{MARNet architecture}\label{sec:marnetab}

\begin{figure}[t]
    \centering
    \resizebox{.9\linewidth}{!}{
\begin{tikzpicture}

\definecolor{crimson2143940}{RGB}{214,39,40}
\definecolor{darkgray176}{RGB}{176,176,176}
\definecolor{darkorange25512714}{RGB}{255,127,14}
\definecolor{forestgreen4416044}{RGB}{44,160,44}
\definecolor{lightgray204}{RGB}{204,204,204}
\definecolor{mediumpurple148103189}{RGB}{148,103,189}
\definecolor{orchid227119194}{RGB}{227,119,194}
\definecolor{sienna1408675}{RGB}{140,86,75}
\definecolor{steelblue31119180}{RGB}{31,119,180}

\begin{axis}[
legend cell align={left},
legend style={
  fill opacity=0.8,
  draw opacity=1,
  text opacity=1,
  at={(0.97,0.03)},
  anchor=south east,
  draw=lightgray204
},
tick align=outside,
tick pos=left,
x grid style={darkgray176},
xlabel={Epoch},
xmin=-95, xmax=1995,
xtick style={color=black},
y grid style={darkgray176},
ylabel={PSNR},
ymin=6.27296341331111, ymax=62.79359073265,
ytick style={color=black}
]
\addplot [semithick, steelblue31119180]
table {%
0 33.7817499286276
100 52.5826122727756
200 54.0244127290941
300 56.3410355067715
400 57.0928308764372
500 57.6742842726591
600 57.478148470396
700 58.0159954213117
800 58.0691092124494
900 57.9246012756522
1000 58.3899900685178
1100 57.1219890425984
1200 57.8528728532225
1300 57.8949604644686
1400 58.1612349226416
1500 57.9308512906487
1600 57.8745969878208
1700 57.6601894455099
1800 57.2764230110651
1900 56.6927821437416
};
\addlegendentry{ResNet}
\addplot [semithick, darkorange25512714]
table {%
0 9.01085236722181
100 44.9498202616882
200 46.9748395564191
300 48.8992092840904
400 50.9240415950169
500 50.7167732773126
600 51.0436220968992
700 51.8695965685544
800 52.8985608580515
900 53.2522998950695
1000 53.869656597453
1100 54.0360269791212
1200 54.0703009008091
1300 54.0922971043632
1400 54.1492111120811
1500 54.0653225610509
1600 53.8689025452233
1700 53.20001772006
1800 53.8082399421568
1900 52.9586496652449
};
\addlegendentry{MARNet1}
\addplot [semithick, forestgreen4416044]
table {%
0 17.5462170715991
100 45.1317032515623
200 49.2383573882371
300 49.6848299703626
400 50.7173360759514
500 51.3454266123482
600 51.4403609295687
700 51.9696982505104
800 51.9032644379829
900 52.6088486556026
1000 52.6285415516663
1100 52.7919763225569
1200 53.0351536160729
1300 53.0071861147634
1400 53.0377780622846
1500 53.085470270117
1600 52.9304549526081
1700 53.1608590786767
1800 53.0452431942419
1900 53.2888033958126
};
\addlegendentry{MARNet2}
\addplot [semithick, crimson2143940]
table {%
0 27.4335082270614
100 48.8905108804201
200 51.0744681806363
300 51.5475734103191
400 51.6944578911483
500 51.9083945426906
600 52.4891042707429
700 53.4841539345389
800 54.1626687739801
900 54.5262379799018
1000 54.9156565619586
1100 55.0918313982166
1200 54.8389870386747
1300 54.9734395522562
1400 54.8550549206424
1500 54.6809793398804
1600 54.3950536428014
1700 53.8065732579566
1800 54.4733724878888
1900 54.4832575515851
};
\addlegendentry{MARNet3}
\addplot [semithick, mediumpurple148103189]
table {%
0 9.35843347450921
100 39.3542010217262
200 52.0541569650499
300 52.6195451776861
400 52.5363051526052
500 52.6435298113322
600 53.2871711448341
700 53.9161521857139
800 54.8656721519473
900 55.6190265926575
1000 56.1153347576524
1100 56.9138848968468
1200 57.1902023818146
1300 57.2064793834024
1400 56.691615479081
1500 57.5011291333751
1600 57.709162536409
1700 57.8160673068083
1800 57.5601136011816
1900 58.199217844848
};
\addlegendentry{MARNet4}
\addplot [semithick, sienna1408675]
table {%
0 8.84208283691742
100 37.6300489745866
200 47.4233575122392
300 48.5449724121305
400 48.2020468601917
500 48.5135219300413
600 49.4031942683976
700 50.2603228949015
800 49.6128007897479
900 51.8012707568893
1000 52.28726786983
1100 52.2772826341238
1200 53.1949833401635
1300 53.507790578413
1400 53.1728122034807
1500 54.0880441836933
1600 54.1721096796746
1700 54.3702142429976
1800 54.7491513256744
1900 54.7046609775991
};
\addlegendentry{MARNet5}
\addplot [semithick, orchid227119194]
table {%
0 17.3418339353395
100 48.18841277326
200 52.2507504320215
300 54.8066871120689
400 56.2281638677126
500 56.3011690113645
600 58.518155826485
700 58.9927442438856
800 59.0392299915648
900 58.8680481219232
1000 58.8651212153961
1100 59.0449356611961
1200 59.2296453872235
1300 59.6479754535194
1400 60.0203600034203
1500 60.1508018007259
1600 59.7704404647263
1700 60.2244713090437
1800 59.704752431305
1900 60.1134475285848
};
\addlegendentry{MARNet6}
\end{axis}

\end{tikzpicture}}
    \caption{For each MARNet architecture proposed in Subsection \ref{sec:marnetab}, the curve describes the evolution of the PSNR on the validation set over 2000 epochs for image denoising tasks. A comparison with a simple residual network is also included. We observe that MARNet6, which corresponds to the architecture presented in Figure \ref{fig:marnet}, provides the best performance.}
    \label{fig:marnet_ablation}
\end{figure}
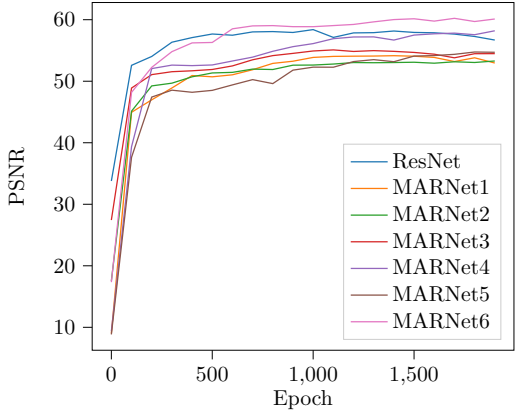

The architecture of the MARNet has been extensively investigated. In particular, we evaluated six different configurations concerning the inputs to the residual blocks and the multi-head attention module. For the residual blocks, we tested whether the grayish concatenation in Figure \ref{fig:marnet}a is used (MARNet4-MARNet6) or not (MARNet1-MARNet3). Within each of these two groups, the variations in the architecture are based on the feature extraction applied before the multi-head attention module, depicted in brown in Figure \ref{fig:marnet}a. This involves determining whether feature extraction is applied to the input data (MARNet1, MARNet4), the PAN image (MARNet2, MARNet5), or both (MARNet3, MARNet6).

Since the proximity operator \eqref{eq:prox} can be interpreted as the solution to a denoising problem with an arbitrary regularization term, we evaluated the performance of all architectures on denoising tasks. We selected two sets from DIV2K \cite{agustsson2017workshops}, with 13 images for training and 6 for validation purposes. Each architecture was trained using the MSE loss function for 2000 epochs. For each configuration, Figure \ref{fig:marnet_ablation} illustrates the evolution of the PSNR on the validation set throughout all epochs. Additionally, we compare these results with those obtained using a simple residual network replacing the proximity operator. We observe that MARNet6, which corresponds to the architecture presented in Figure \ref{fig:marnet}, provides the best performance.

\subsection{Post-processing}

For the post-processing module, we evaluated the performance of a residual-based architecture, a CNN-based architecture, and the proposed MARNet. Additionally, we compared these with the results obtained without any post-processing. All four configurations were trained on the PRISMA dataset over 1000 epochs and using the loss functions presented in \eqref{eq:lossFinal}. Table \ref{tab:ablation-post} displays the quantitative metrics for each configuration. The results demonstrate that the best performance is achieved when using the MARNet for post-processing.

\begin{table}[t]
    \centering
    \caption{Quantitative metrics for different post-processing configurations on the PRISMA dataset. The best performance (highlighted in bold) is achieved when the MARNet is used as post-processing.}
    \begin{tabular}{l|ccccc}
         &  ERGAS$\downarrow$&  PSNR$\uparrow$&  SSIM$\uparrow$&  Q$2^n\uparrow$& SAM$\downarrow$\\
         \hline
         No post& 26,23& 41,53& 0,953& 0,831& 2,74\\
         ResNet&  44,33&  37,92&  0,935&  0,712 & 5,42\\
 CNN& 26,90& 41,29& 0,952& 0,818&2,90\\
 MARNet& \textbf{25,77}& \textbf{41,72}& \textbf{0,956}& \textbf{0,832}&\textbf{2,71}\\
 \hline
    \end{tabular}
    \label{tab:ablation-post}
\end{table}

\subsection{Robustness to sampling factor}

To assess the robustness of the proposed method with respect to the sampling factor, we used two images from the Pelican dataset \cite{duran2017survey}, provided by the French Space Agency (CNES). This dataset consists of 4-band images (blue, green, red, and near-infrared) at 10 cm collected by an airborne platform. From them, we built reference spectral channels at 30 cm by MTF filtering and subsampling, used to evaluate the fused products in terms of quantitative metrics. PAN images were generated as a linear combination of the original spectral bands. We cropped the reference and PAN images into patches of size $144\times 144$ pixels, which were subsequently divided into training and validation sets with ratios of 80$\%$ and 20$\%$, respectively. The low-resolution MS data were produced using four different sampling factors $s\in\{4, 12, 24, 36\}$. 

\begin{figure*}[t]
    \centering
    \begin{tabular}{c@{\hskip 0.2em}c@{\hskip 0.2em}c@{\hskip 0.2em}c}
         \includegraphics[width=0.2\linewidth]{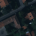} &
         \includegraphics[width=0.2\linewidth]{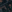} &
         \includegraphics[width=0.2\linewidth]{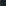} &
         \includegraphics[width=0.2\linewidth]{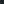}
         \\
        MS ($s=4$) & MS ($s=12$) &  MS ($s=24$)& MS ($s=36$)
         \\
         \includegraphics[width=0.2\linewidth]{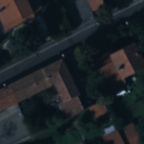} & \includegraphics[width=0.2\linewidth]{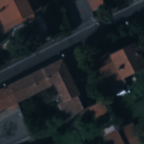} & \includegraphics[width=0.2\linewidth]{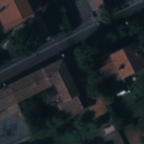} & \includegraphics[width=0.2\linewidth]{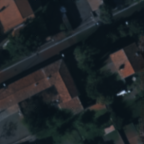}
         \\
        Fused ($s=4$)& Fused ($s=12$) &  Fused ($s=24$) & Fused ($s=36$)
         \\
        
    \end{tabular}
    \caption{Visual comparison of the low-resolution MS data and the fused images produced by the proposed method on the Pelican dataset for different sampling factors. The results demonstrate geometric and spectral consistency, even in highly challenging scenarios such as $s=24$ and $s=36$.}

    \label{fig:abltaion-sampling}
\end{figure*}

Table \ref{tab:ablation-sampling} provides the quantitative results for each sampling factor, and Figure \ref{fig:abltaion-sampling} displays the MS data and the fused images. We observe that, even in highly challenging scenarios such as $s=24$ and $s=36$, our method is able to effectively recover the geometry of the scene while ensuring spectral consistency.

\begin{table}[t]
    \centering
    \caption{Quantitative metrics obtained by the proposed fusion method on the Pelican dataset for different sampling factors.}
    \begin{tabular}{c|ccccc}
         s &  ERGAS$\downarrow$&  PSNR$\uparrow$&  SSIM$\uparrow$&  Q$2^n\uparrow$& SAM$\downarrow$\\
         \hline
         4 &  13,93 &  47,70 &  0,994 &  0,978 & 1,890\\
         12 &  35,34 &  39,63 &  0,979 &  0,914 & 3,211\\
         24 &  57,12 &  35,37 &  0,964 &  0,850 & 4,605 \\
         36 &  68,12 &  33,91 &  0,961 &  0,844 & 5,509 \\
         \hline
    \end{tabular}
    \label{tab:ablation-sampling}
\end{table}

\begin{table}[t]
    \centering
    \caption{Quantitative metrics obtained by the proposed fusion method on the Pelican dataset for different noise levels.}
    \begin{tabular}{c|ccccc}
         $\sigma$ &  ERGAS$\downarrow$&  PSNR$\uparrow$&  SSIM$\uparrow$&  Q$2^n\uparrow$& SAM$\downarrow$\\
         \hline
         0 &  13,93 &  47,70 &  0,994  &  0,978 & 1,890\\
         5 &  30,15 &  40,95 &  0,983  &  0,911 & 2,854 \\
         10 &  36,02 &  39,39 &  0,981  &  0,896 & 3,413 \\
         25 &  58,06 &  35,31 &  0,953  &  0,794 & 4,780 \\
         50 &  68,45 &  33,92 &  0,958  &  0,804 & 5,952 \\
         \hline
    \end{tabular}
    \label{tab:ablation-noise}
\end{table}

\begin{figure*}[t]
    \centering
    \begin{tabular}{c@{\hskip 0.2em}c@{\hskip 0.2em}c@{\hskip 0.2em}c}
         \includegraphics[width=0.2\linewidth]{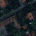} &
         \includegraphics[width=0.2\linewidth]{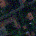} & 
         \includegraphics[width=0.2\linewidth]{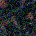} & 
         \includegraphics[width=0.2\linewidth]{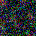}
         \\
        MS ($\sigma=5$)& MS ($\sigma=10$)&  MS ($\sigma=25$)& MS ($\sigma=50$)\\
         \includegraphics[width=0.2\linewidth]{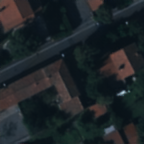} & \includegraphics[width=0.2\linewidth]{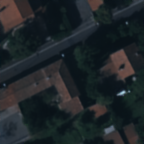} & \includegraphics[width=0.2\linewidth]{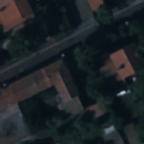} & \includegraphics[width=0.2\linewidth]{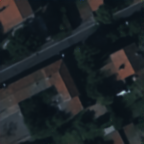}
         \\
        Fused ($\sigma=5$)& Fused ($\sigma=10$)&  Fused ($\sigma=25$)& Fused ($\sigma=50$)\\
        
    \end{tabular}
    \caption{Visual comparison of the noisy low-resolution MS data and the fused images produced by the proposed method on the Pelican dataset for different noise levels. The sampling factor used in these experiments was $s=4$. The results demonstrate geometric and spectral consistency, even for $\sigma=50$.}
    \label{fig:abltaion-noise}
\end{figure*}

\subsection{Robustness to noise}

Finally, we evaluate the robustness of our fusion method to noise. As in the previous subsection, we use the Pelican dataset and introduce noise into the low-resolution MS data. Specifically, we added different noise realizations following a normal distribution with standard deviations of $\sigma\in\{5, 10, 25, 50\}$. The sampling factor used in these experiments was $s=4$.

Table \ref{tab:ablation-noise} shows the quantitative metrics for each noise level, while Figure \ref{fig:abltaion-noise} displays the noisy MS data alongside the fused products. The results demonstrate the robustness of our approach to noise, showing that the fused images maintain high spatial and spectral quality even for $\sigma=50$.

\section{Conclusions}\label{sec:conclusion}

In this paper, we proposed a novel model-based deep unfolded method for satellite image fusion. We unrolled the primal-dual optimization scheme derived from minimizing a variational formulation, which includes a classic observation model for MS/HS data, a high-frequency injection constraint based on the PAN image, and an arbitrary convex prior. For the unfolding stage, we introduced upsampling and downsampling layers that use the information encoded in the PAN image through residual networks and can be adapted to arbitrary sampling factors. The core of our method is the MARNet module, which replaces the proximity operator in the primal-dual scheme and combines multi-head attention mechanisms with residual learning to leverage image self-similarities through nonlocal operations defined in terms of patches. The multiple head attentions are designed to capture spatial and spectral information across both PAN and MS/HS data. Additionally, we incorporated a post-processing module based on the MARNet architecture. The ablation study demonstrates that this post-processing significantly enhances the quality of the fused images.

To evaluate the performance of the proposed method and compare it with state-of-the-art fusion approaches, we conducted experiments on various datasets, testing the generalization capabilities of competing techniques. We used data from PRISMA, Quickbird, and WorldView2 satellites, each with different sensor configurations and varying spatial and spectral resolutions. Our proposal achieved the best results on PRISMA and WorldView2, and was the only method to rank in the top three for all metrics on Quickbird, where other approaches exhibited inconsistent performance. Furthermore, we demonstrated the robustness of our model against variations in the sampling factor and noise. Even in challenging scenarios, our fused images maintained high spatial and spectral quality.

Some advantages of the proposed method include: its efficient and highly interpretable architecture due to its model-based nature; robust adaptability to different sensor configurations and varying spatial and spectral resolutions; ease of adaption to MS+HS image fusion; capability to be trained with an arbitrary sampling factor while still providing results with accurate spatial and spectral consistency; and robustness to different noise levels. Despite these strengths, future work is needed to: investigate more efficient strategies for computing attention mechanisms while extracting information from patches; develop new energy terms leveraging the spectral information encoded in the MS/HS modalities; and introduce adaptive convolutions in the context of unfolding.

\backmatter





\bmhead{Acknowledgements}
This work was funded by the European Union NextGenerationEU/PRTR via MaLiSat project TED2021-132644B-I00, and also by MCIN/AEI/10.13039/501100011033 and “ERDF A way of making Europe” through European Union under Grant PID2021-125711OB-I00.  We are also grateful for the funding provided by the Conselleria de Fons Europeus, Universitat i Cultura (GOIB)  FPU2023-004-C. The authors gratefully acknowledge the computer resources at Artemisa, funded by the EU ERDF and Comunitat Valenciana and the technical support provided by IFIC (CSIC-UV). We are grateful to the Italian Space Agency for giving us access to the PRISMA database.

\section*{Declarations}
\begin{itemize}
\item All of the codes used in this work are available in a public repository.
\item We are not authorized to share the data used; however, we have specified how to request it from the owners' platforms.
\end{itemize}


%
%
%
%


\bibliography{sn-biblio}

\end{document}